\documentclass[draftcls,onecolumn]{IEEEtran}
\usepackage{graphicx}
\usepackage{amsmath}
\usepackage{multicol}
\usepackage{multirow}
\usepackage{stfloats}
\usepackage{booktabs}
\usepackage{mathrsfs}
\usepackage{enumerate}
\usepackage{amssymb}
\usepackage{algorithm}
\usepackage{algpseudocode}

\usepackage{array}
\usepackage{underscore}
\usepackage{url}

\usepackage{hyperref}
\hypersetup{hidelinks}
\usepackage{flushend}

\usepackage[square, comma, sort&compress, numbers]{natbib}
\usepackage{graphicx}
\usepackage{epstopdf}
\usepackage{caption}
\usepackage{diagbox}

\usepackage{caption}
\usepackage{subfigure}
\usepackage{color}

\usepackage[outerbars,color]{changebar}
\ifx\pdfoutput\undefined
\else\ifnum\pdfoutput>0
  \usepackage{pdfcolmk}
\fi\fi
\cbcolor{red}
\usepackage{fancyhdr}

\makeatletter
\newcommand{\Rmnum}[1]{\expandafter\@slowromancap\romannumeral #1@}
\makeatother

\usepackage{array}  \newcommand{\PreserveBackslash}[1]{\let\temp=\\#1\let\\=\temp}  \newcolumntype{C}[1]{>{\PreserveBackslash\centering}p{#1}}  \newcolumntype{R}[1]{>{\PreserveBackslash\raggedleft}p{#1}}  \newcolumntype{L}[1]{>{\PreserveBackslash\raggedright}p{#1}}

\ifCLASSINFOpdf
\else
\fi
\begin{document}

\title{\begin{Huge}Fair Scheduling in Resonant Beam Charging\\ for IoT Devices\end{Huge}}

\author{Wen~Fang,
Qingqing~Zhang,
Qingwen~Liu\IEEEauthorrefmark{1}, 
Jun~Wu, 
and~Pengfei~Xia	

	\thanks{W.~Fang, Q. Zhang, Q. Liu, J. Wu, and P. Xia, are with College of Electronic and Information Engineering, Tongji University, Shanghai, People's Republic of China, (e-mail: wen.fang@tongji.edu.cn, anne@tongji.edu.cn, qliu@tongji.edu.cn,  wujun@tongji.edu.cn, and pengfei.xia@gmail.com).}%
    \thanks{* Corresponding author.}
}

\maketitle

\begin{abstract}

    Resonant Beam Charging (RBC) is the Wireless Power Transfer (WPT) technology, which can provide high-power, long-distance, mobile, and safe wireless charging for Internet of Things (IoT) devices. Supporting multiple IoT devices charging simultaneously is a significant feature of the RBC system. To optimize the multi-user charging performance, the transmitting power should be scheduled for charging all IoT devices simultaneously. In order to keep all IoT devices working as long as possible for fairness, we propose the First Access First Charge (FAFC) scheduling algorithm. Then, we formulate the scheduling parameters quantitatively for algorithm implementation. Finally, we analyze the performance of FAFC scheduling algorithm considering the impacts of the receiver number, the transmitting power and the charging time. Based on the analysis, we summarize the methods of improving the WPT performance for multiple IoT devices, which include limiting the receiver number, increasing the transmitting power, prolonging the charging time and improving the single-user's charging efficiency. The FAFC scheduling algorithm design and analysis provide a fair WPT solution for the multi-user RBC system.

\end{abstract}


\IEEEpeerreviewmaketitle

\section{Introduction}\label{Section1}

    Internet of things (IoT) aims at connecting ubiquitous devices to Internet, and it has become an  important driving force for information technology innovation \cite{Campbell2016, wu2014cognitive}. Power supply is crucial to support high-performance computation and communication in IoT applications. However, the power sustainability is the well-known headache for battery-assisted IoT devices. There are two ways to extend the battery endurance: 1) increasing the battery capacity, and 2) improving the battery charging method.

    Increasing the battery capacity faces the challenges, e.g., safety, weight, cost, recycling and so on \cite{scrosati2010lithium}. On the other hand, for the wired charging method, carrying a power cord and looking for a power outlet cause inconvenience for users. Hence, wireless power transfer (WPT), as known as wireless charging, becomes an attractive solution to improve battery endurance \cite{lu2015wireless}.

    Being able to transmit Watt-level power over meter-level distance safely, resonant beam charging (RBC), as known as distributed laser charging (DLC), was presented in \cite{liu2016dlc}. Comparing with the other wireless charging methods, e.g., inductive coupling, magnetic resonance coupling, radio frequency, and so on, RBC appears to be more suitable for mobile IoT devices \cite{ho2011comparative, kurs2007wireless, costanzo2014electromagnetic}. In the RBC system, the wireless power transmitter and receiver are separated in space. Without specific aiming or tracking, the resonant beam can be generated as long as the receiver is in the line of sight (LOS) of the transmitter, and multiple resonant beams can be generated from one transmitter to multiple receivers \cite{liu2016dlc}. Thus, RBC can charge multi-device simultaneously like Wi-Fi communications.

    Due to various types and working status of the RBC receivers (e.g., IoT devices), the battery remaining capacity percentage (i.e., state of charging, SOC) and discharging status may be diverse \cite{SOC2006}. Therefore, all receivers' charging status (e.g., preferred charging power, charging time) may be different as well. However, in the multi-user RBC application scenario, such as the wireless sensor network discussed in \cite{lewis2004wireless}, if one receiver's battery exhausts, the network system may break down \cite{yu2015malware, ding2013sensing}. Thus, in order to keep all IoT devices in the system working as long as possible, it is necessary to study the scheduling method for the multi-user RBC system.

    \begin{figure}[!t]
	\centering
    \includegraphics[scale=0.4]{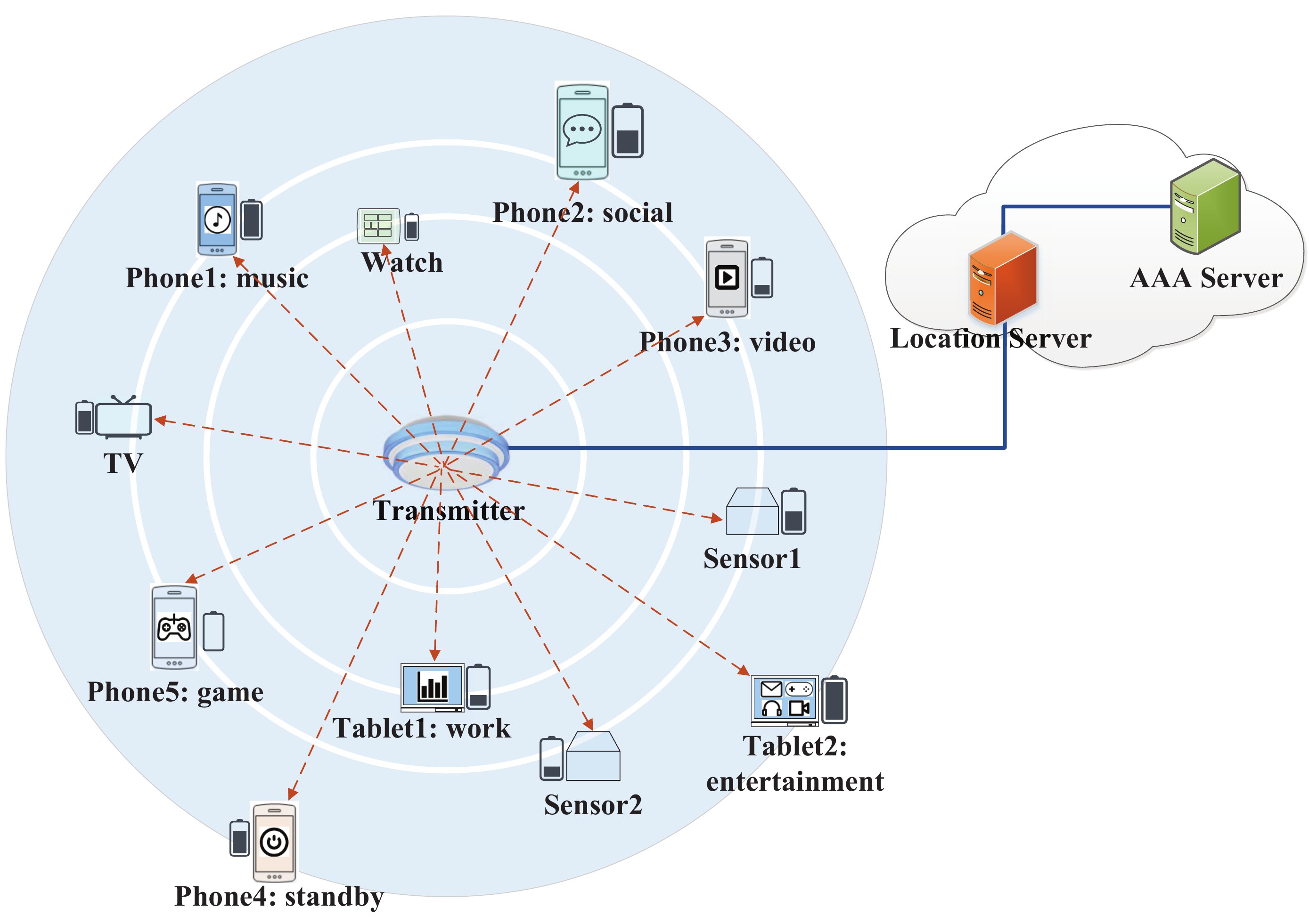}
	\caption{Multi-user RBC Application Scenario}
    \label{complex-scenarios}
    \end{figure}

     The system consisting of a RBC transmitter and multiple receivers is defined as the multi-user RBC system. Fig. \ref{complex-scenarios} shows a multi-user RBC application scenario. In Fig. \ref{complex-scenarios}, all receivers, including mobile phones, watches, tablets and sensors, are operated and charged at the same time. Since the receivers' batteries and the working status are different, the receivers have different power charging and discharging status. Thus, each receiver SOC is different. If the transmitting power can not satisfy the charging power requirements of all receivers simultaneously, several receivers' batteries may exhaust due to ``starvation". Therefore, it is desired to design the scheduling algorithm to keep all receivers working as long as possible for fairness in the multi-user RBC system.


    The contributions of this paper include: 1) We propose the First Access First Charge (FAFC) scheduling algorithm for WPT in the multi-user RBC system, which can keep all receivers working as long as possible for fairness. 2) We obtain the closed-form formulas of the parameters for the FAFC scheduling algorithm implementation based on the quantitative analysis. 3) We analyze the performance of the FAFC scheduling algorithm, and find the features of the algorithm as:
    \begin{itemize}
      \item When the transmitting power is fixed, there exists a threshold for the number of receivers being charged simultaneously in order to avoid the system running out of power. For example, the maximum receiver number threshold is about 35 when the transmitting power is 20W.
      \item If the transmitting power can satisfy the consumed power of all receivers, the charging time should be prolonged to extend the working time of all receivers.
      \item Regardless of the receiver number and the charging time, the system operational duration increases when increasing the transmitting power or improving the single-user's charging efficiency.
    \end{itemize}

    In the rest of this paper, we will depict the multi-user RBC system in Section II. In Section III, we will design the FAFC scheduling algorithm and present its execution flow. In Section IV, we will propose the quantitative formulation methods for algorithm implementation. In Section V, we will analyze the performance of the FAFC scheduling algorithm by MATLAB simulation.

\section{Multi-User RBC System}\label{Section2}
\begin{figure}[!t]
	\centering
    \includegraphics[scale=0.5]{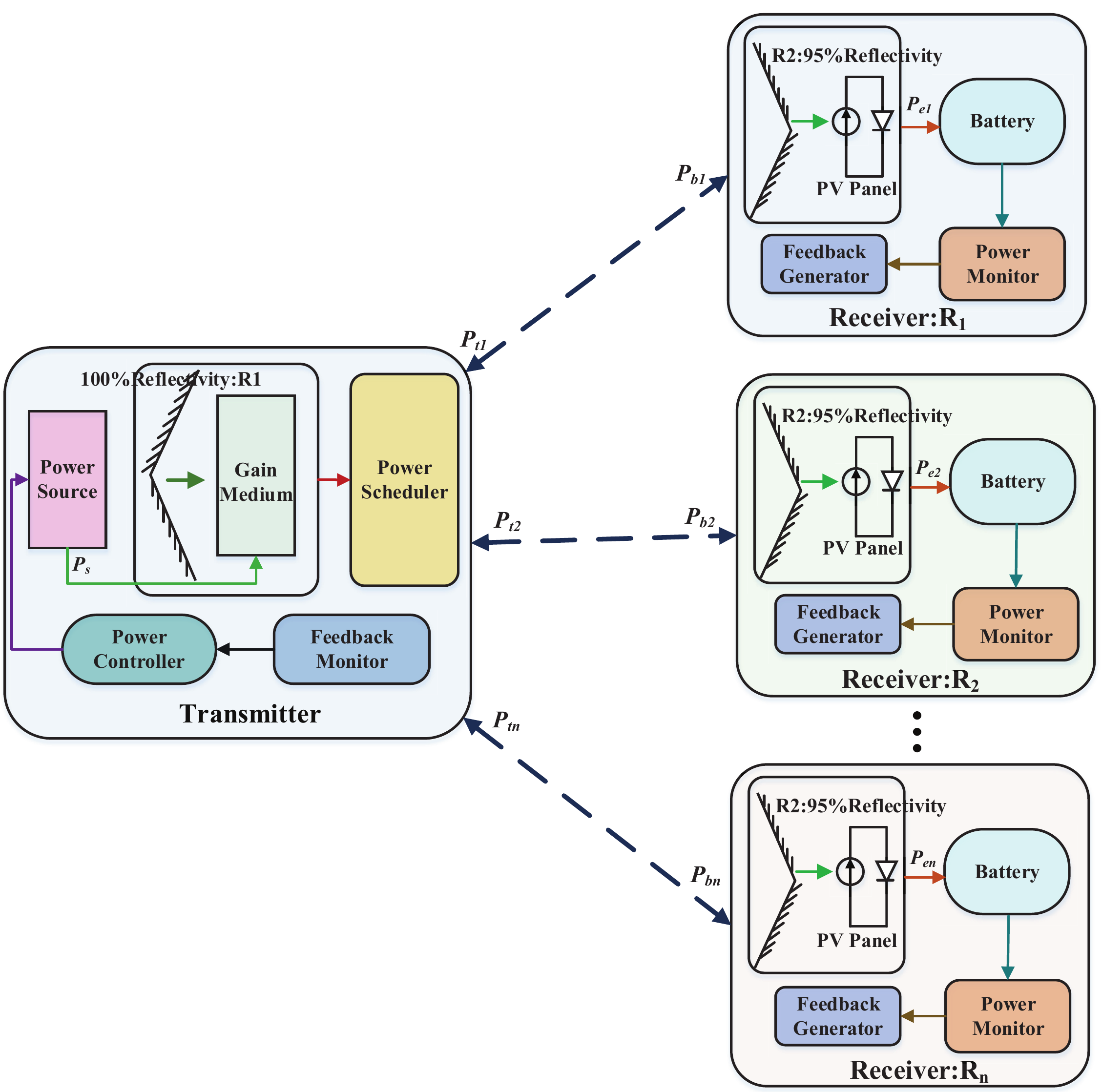}
	\caption{Multi-user RBC System}
    \label{RBC}
    \end{figure}

    The RBC system can transmit Watt-level power over meter-level distance while guaranteeing the mobile and safe charging for IoT devices. In addition, multiple receivers can be charged by a transmitter simultaneously \cite{liu2016dlc}. The system with a transmitter and multiple receivers is called the multi-user RBC system, as shown in Fig. \ref{RBC}.

    In Fig. \ref{RBC}, the multi-user RBC system consists of one transmitter and multiple receivers. The transmitter contains a power source, a retro-reflector R1 with 100\% reflectivity, a gain medium, a power scheduler, a feedback monitor and a power controller. The power source provides electrical power $P_s$ for the gain medium under the control of the power controller. After stimulating the gain medium, the electrical power $P_s$ is converted to the beam power at the transmitter (i.e., transmitting power) $P_t$. The power scheduler controls the arrangement of receiver queue and the transmitting power distribution. The feedback monitor is responsible for receiving and processing feedback information, and informing the power controller with the battery preferred charging information.

    A retro-reflector R2 with 95\% reflectivity, a photovoltaic (PV) panel, a battery, a power monitor and a feedback generator are included in the receiver $R_i$. The power of resonant beam sending to R2 $P_b$ can be partially converted into the receiver output electrical power $P_e$ by using PV panel. The power monitor tracks the battery status of charging and sends information to the feedback generator. Then the generator feeds the information back to the transmitter. The symbols in Fig. \ref{RBC} are listed in Table \ref{mulparameter}.

    The wireless power transfer processes of the multi-user RBC system include: electricity-to-beam conversion, transmitting power scheduling, resonant beam transmission,  beam-to-electricity conversion, charging status feedback, and feedback information processing.

    From Fig. \ref{RBC}, the receivers form $R_{1}$ to $R_n$ have established charging connections with the transmitter. When the fixed transmitting power is less than the preferred charging power of all receivers, the transmitting power can not meet all receivers' charging requirements at the same time. Therefore, the scheduling method should be adopted to control the transmitting power scheduling for charging the receivers.

    \begin{table}[!t]
    \setlength{\abovecaptionskip}{0pt}
    \setlength{\belowcaptionskip}{-3pt}
    \centering
        \caption{Multi-user RBC System Symbols}
    \begin{tabular}{C{1.5cm} C{6.0cm}}
    \hline
     \textbf{Symbol} & \textbf{Parameter}  \\
    \hline
    \bfseries{$P_s$} & {Source power} \\
    \bfseries{$P_t$} & {Beam power, i.e., transmitting power} \\
    \bfseries{$P_b$} & {Receiver beam power } \\
    \bfseries{$P_e$} & {Receiver output electrical power} \\
    \bfseries{$R_i$} & {Serial number of the i-th receiver} \\
    \hline
    \label{mulparameter}
    \end{tabular}
    \end{table}

     We will present the first access first charge (FAFC) scheduling algorithm to solve the scheduling problem for WPT in the multi-user RBC system. The aim of FAFC scheduling algorithm is depicted as:

    \begin{itemize}
      \item The receiver queue is:
      \center
      $R$=\{$R_i$ $\mid$ the $i$-th receiver which has accessed to the transmitter for charging\}.
    \end{itemize}

    In FAFC scheduling process, all the receivers should be kept working as long as possible for fairness, i.e., the SOC (i.e., remaining capacity percentage) $R_{soc}$ of any receiver $R_i$ is greater than 0, that is:
    \begin{equation}\label{purpose}
    \centering
    R_{soc}(R_i)>0,\ \ \ \ (\forall R_i \in R).
    \end{equation}

    In the next section, to satisfy the simultaneous charging needs in the multi-user RBC system, we will specify the FAFC scheduling algorithm in detail.

\section{FAFC Algorithm Design}\label{Section2}

    We will focus here on the multi-user RBC scenario where the working time of all devices is supposed to be maintained as long as possible. Since the receivers' battery SOC and discharging status are different, their charging status is varying. To maximize the utilization of the transmitting power and meet the receivers' charging power requirements, the transmitting power distribution should be controlled and the receivers should be charged with their preferred charging power. To solve these problems, we will study the FAFC scheduling algorithm in this section.
\subsection{Design Ideas}\label{}
    \begin{figure}[!t]
	\centering
    \includegraphics[scale=0.6]{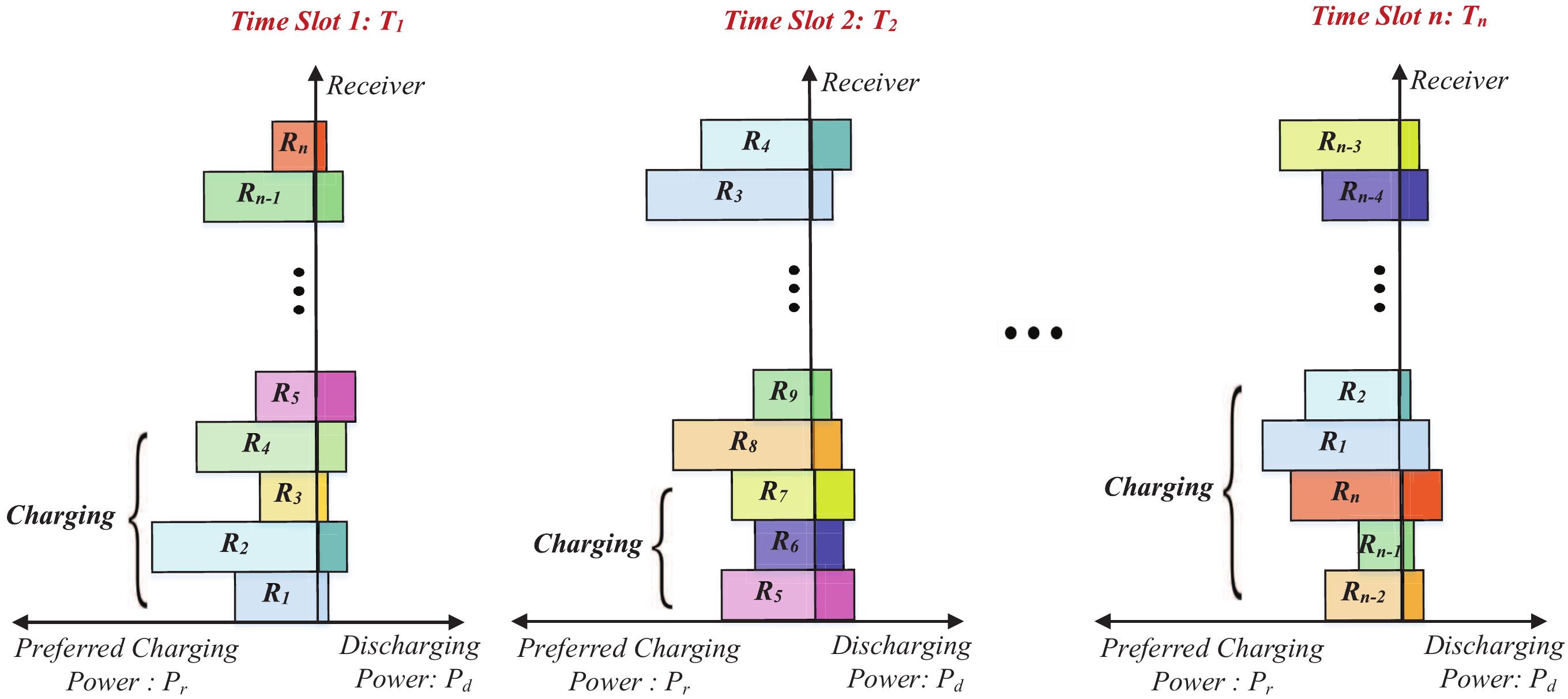}
	\caption{FAFC Schematic Diagram}
    \label{FAFCtime}
    \end{figure}
    The design ideas of the FAFC scheduling algorithm include: 1) The charging time is divided into several identical charging time slots. 2) The receivers accessed to the transmitter are queued up according to their accessing time chronologically. 3) The multiple receivers in the queue's head can be charged with their preferred charging power during a charging time slot,  while the other receivers keep waiting. 4) All the receivers discharge with different power according to their working status in each charging time slot. 5) At the end of the time slot, all the receivers update their SOC, and the receivers which have been charged are queued to the tail of the receiver queue. The schematic diagram of the FAFC scheduling algorithm is shown in Fig. \ref{FAFCtime}.

    Fig. \ref{FAFCtime} shows the FAFC scheduling principles with $n$ receivers from $R_1$ to $R_n$. Each rectangle denotes a RBC receiver $R_i$ with its preferred charging power and discharging power. The left horizontal axis represents the preferred charging power $P_r$, the right one represents the discharging power $P_d$, and the ordinate axis represents the receivers. The preferred charging power (the length of the left rectangle) and discharging power (the length of the right rectangle) of each receiver are different.
    \begin{figure}[!t]
	\centering
    \includegraphics[scale=0.6]{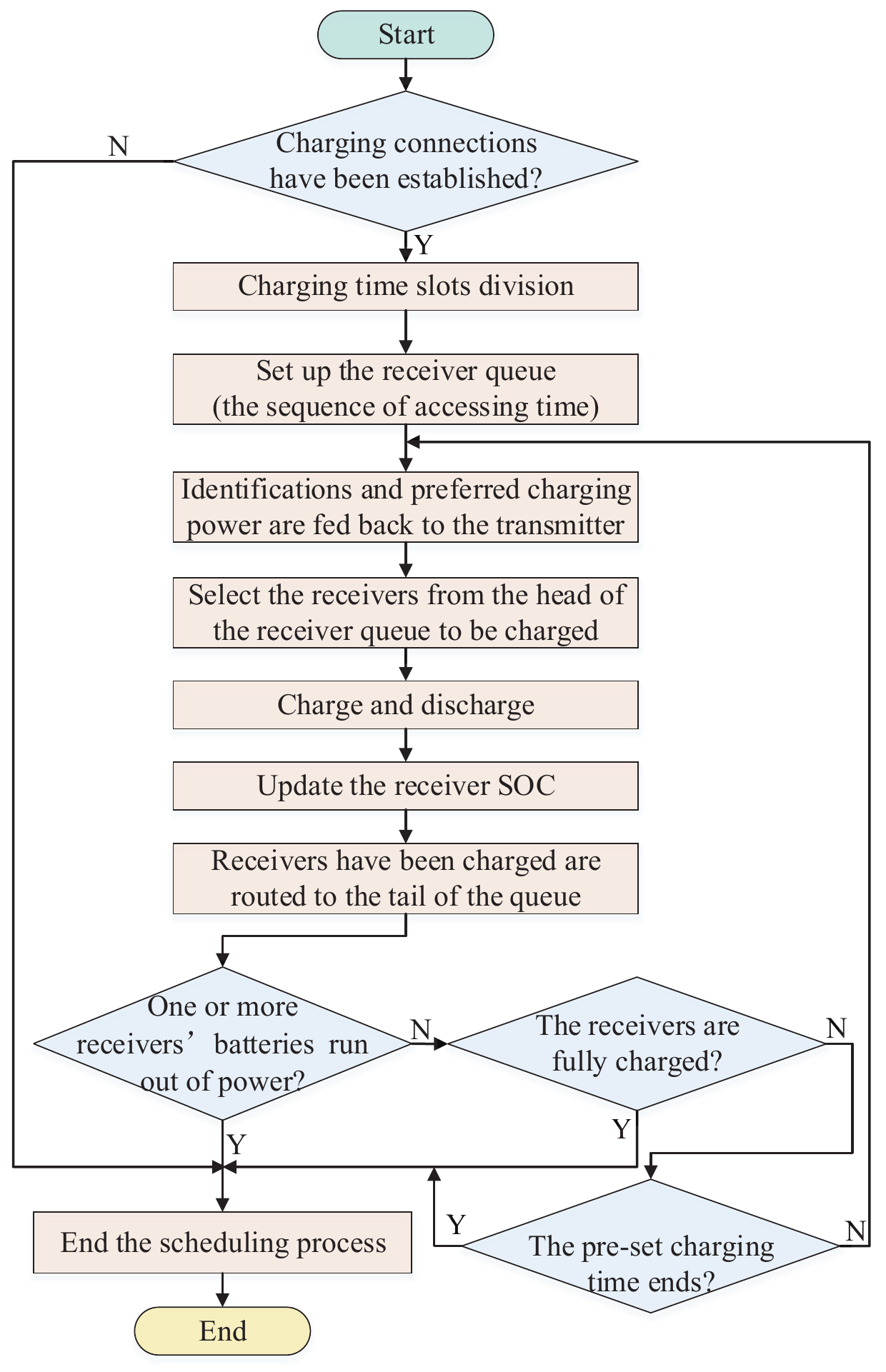}
	\caption{FAFC Scheduling Algorithm Execution Flow}
    \label{FAFCAflowchart}
    \end{figure}
    In Fig. \ref{FAFCtime}, the receivers in the bracket ``charging" are chosen to be charged, which will be routed to the tail of the receiver queue when the charging time slot ends. For example, when the time slot $T_1$ ends, the receivers $R_1$, $R_2$, $R_3$, $R_4$ are placed to the tail of receiver queue. All receivers from $R_1$ to $R_n$ are discharging during the whole scheduling process. The execution flow of the FAFC scheduling algorithm will be discussed in the next subsection.

\subsection{Execution Flow}\label{}
    The execution flow of the FAFC scheduling algorithm is depicted in Fig. \ref{FAFCAflowchart}:

    1) Access judgement: if the charging connections between the transmitter and the receivers have not been established, the scheduling process ends. Otherwise, go to 2).

    2) Time division: the charging time is divided into multiple small charging time slots.

    3) Queuing: according to the time sequence of the receivers accessing to the transmitter, the transmitter sets up the receiver queue of all accessed receivers chronologically.

    4) Feedback: the receiver identifications (receiver type, serial number, battery type, etc.) and battery preferred charging power are fed back to the transmitter.

    5) Selecting: depending on the relationship between the transmitting power and the receivers' preferred charging power, one or more receivers at the head of the receiver queue are chose to be charged.

    6) Charging and discharging: the transmitter schedules transmitting power to charge selected receivers with their preferred charging power. All receivers discharge with different power according to their working status.

    7) Updating: all receivers update their SOC according to the charging and discharging energy.

    8) Rearrangement: the receivers which have been charged are queued to the tail of receiver queue.

    9) End judgement: the scheduling process ends when one of the following three conditions is met: a) one or more receivers' batteries run out of power, b) all receivers are fully charged, c) the pre-set charging time (for example 1hour, 2hours, etc.) ends. Otherwise, turn to 4).

\begin{figure}[!t]
	\centering
    \includegraphics[scale=0.63]{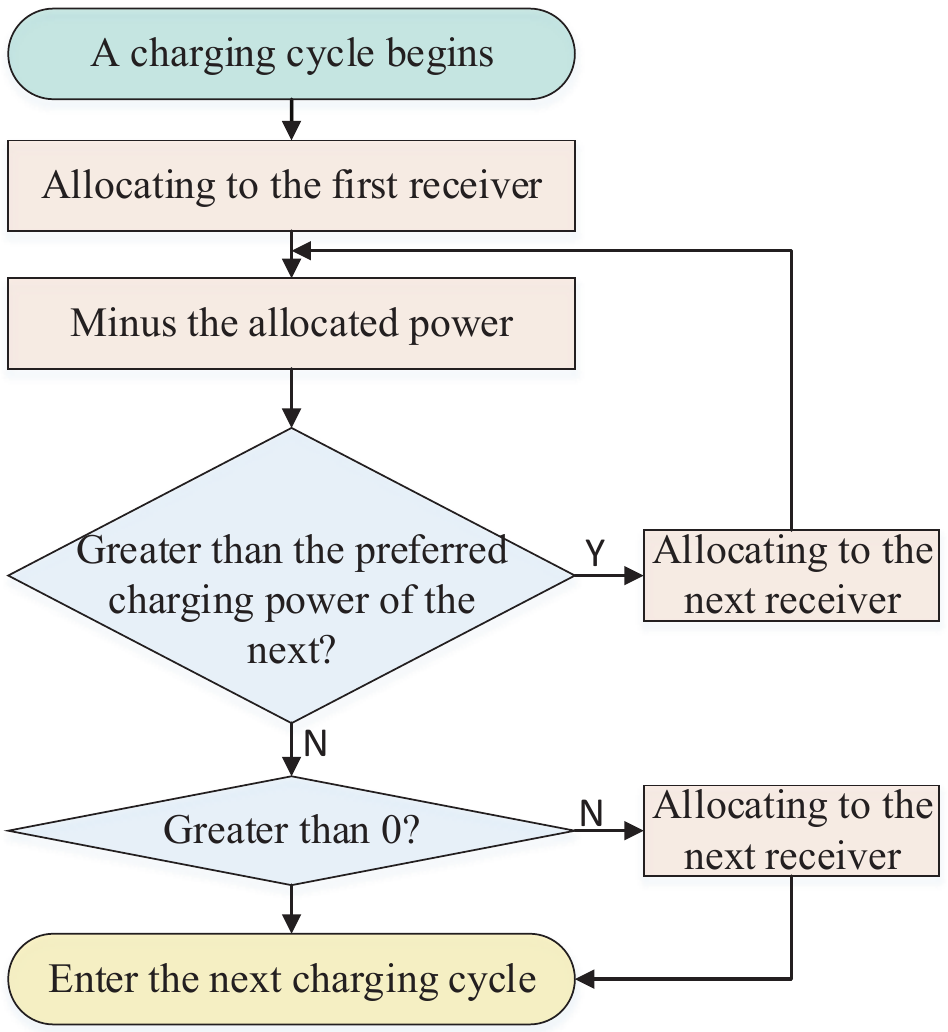}
	\caption{Charging Power Allocation Process}
    \label{chargingcycleallocate}
    \end{figure}

    In addition, during a charging time slot, the power scheduler allocates all the transmitting power to the receivers. Before the transmitting power is allocated, the available transmitting power $P_o$ is equal to the transmitting power $P_t$. The allocating process of the transmitting power $P_t$ is shown in Fig. \ref{chargingcycleallocate}.

    In Fig. \ref{chargingcycleallocate}, $P_{ri}$ is the preferred charging power of the receiver $R_i$, and $P_{ci}$ is the allocated charging power for $R_i$ by the transmitter. For each receiver, $P_{ci}$ is less than or equal to $P_{ri}$. The symbols of the FAFC scheduling algorithm are shown in Table \ref{Parameters of scheme}.

    \begin{table}[!t]
    \centering
        \caption{FAFC Scheduling Algorithm Symbols}
    \begin{tabular}{C{1.5cm} C{6.0cm}}
    \hline
     \textbf{Symbol} & \textbf{Parameter}  \\
    \hline
    \bfseries{$P_o$} & {Transmitter available transmitting power} \\
    \bfseries{$P_r$} & {Battery preferred charging power} \\
    \bfseries{$P_d$} & {Receiver discharging power} \\
     \bfseries{$T_i$} & {Charging time slot} \\
     \bfseries{$P_c$} & {Receiver allocated charging power} \\
    \hline
    \label{Parameters of scheme}
    \end{tabular}
    \end{table}

    From Fig. \ref{chargingcycleallocate}, the transmitting power allocation needs two judging processes: 1) if the $P_o$ is greater than the battery preferred charging power of the next receiver $P_{ri}$, 2) if the $P_o$ is greater than zero Watt. For 1), the remaining transmitting power is allocated to the next receiver according to the preferred value $P_{ri}$, and subtracted the corresponding numerical value from the remaining value. For 2), the remaining transmitting power is all allocated to the next receiver and reduced to zero, then the next charging time slot begins.

    In this section, the FAFC scheduling algorithm was presented. However, to implement the algorithm, various parameters, such as the battery preferred charging power and the discharging power, still need to be specified. In the next section, we will discuss the implementation of the FAFC scheduling algorithm, which includes the quantitative models of the scheduling parameters and the operation pseudo-code of the algorithm.
\section{FAFC Algorithm Implementation}

    In the multi-user RBC system, when the transmitting power is scheduled to keep all receivers working as long as possible for fairness, the algorithm implementation should be presented to quantitatively analyze the charging characteristics. In this section, we will specify the FAFC scheduling algorithm implementation by designing the quantitative models of scheduling parameters. Then, we will illustrate the operation pseudo-code of the algorithm.

\subsection{System Parameters}\label{}

    To quantitatively analyze the scheduling algorithm, we need to specify the following parameters:\\
    1) Charging Efficiency

    In the multi-user RBC system, the charging efficiency ${\eta}_{o}$ of a charging link, which is affected by the electro-optical conversion efficiency ${\eta}_{el}$, the beam transmission efficiency ${\eta}_{lt}$, and the photoelectric conversion efficiency ${\eta}_{le}$ \cite{Qing2017}, can be depicted as:
     \begin{equation}\label{etao}
    {\eta}_{o}={\eta}_{el} {\eta}_{lt} {\eta}_{le}.
    \end{equation}

    The impacting factors on  ${\eta}_{o}$ include: the resonant beam wavelength, the PV panel temperature, the transmission environment (clear, fog, haze), the transmission distance, etc. Since the charging efficiency ${\eta}_{o}$ of a multi-user RBC system is fixed, it has no effect on the performance of the WPT scheduling algorithm. We can assume the parameters for the charging efficiency as:
    \begin{itemize}
      \item The resonant beam wavelength is 810nm.
      \item The electro-optical conversion efficiency ${\eta}_{el}$ is 40\% \cite{810nmtransmitter, zhang2017distributed}.
      \item The transmission efficiency ${\eta}_{lt}$ is 100\% \cite{attenuation}.
      \item The photo-electricity converter is the GaAs-based PV panel working at 25 $^{\circ}$C, and the photoelectric conversion efficiency ${\eta}_{le}$ is 50\% \cite{810nmpv, zhang2017distributed}.
    \end{itemize}

    Therefore, the charging efficiency ${\eta}_{o}$ is 20\% calculated by 40\%$\times$100\%$\times$50\%.\\
    2) Transmitting Power

    Determined by the power source, the source power $P_s$ is fixed during a charging process. Since the electro-optical conversion efficiency ${\eta}_{el}$ is 40\%, the transmitting power $P_t$ is fixed as well, and it is equal to $0.4 \times P_s$. The relationships among $P_s$, $P_t$, $P_b$, $P_e$ specified in Fig. \ref{RBC} are depicted as:
    \begin{equation}\label{transpower}
    \begin{aligned}
    P_{e}=P_b {\eta}_{le} = P_t {\eta}_{lt} {\eta}_{le} = P_s {\eta}_{el} {\eta}_{lt} {\eta}_{le}. &
    \end{aligned}
    \end{equation}
    3) Receiver Specification

    The battery types of the same and different receivers are various, such as Li-ion, Ni-MH, etc. \cite{hussein2011review, park2008universal}. The difference of receiver specification mainly reflects in the battery capacity, the charging current and voltage \cite{Anonymous2008, winter2004batteries}.

     Mobile phone has become one of the most widely used receiving device, while the lithium-ion battery is generally used for mobile phones due to its excellent performance, such as high specific energy, high efficiency and rechargeability \cite{tarascon2001issues}. Therefore, we adopt the mobile phones with the lithium-ion battery, of which the capacity is 1000mAh, the constant charging voltage is 4.2V, the constant charging current is 1A, as the RBC receivers.\\ 
    4) Receiver Number

    In the multi-user RBC system, multiple receivers can be charged with their preferred charging power simultaneously. In the scheduling process, the receiver number $N_{r}$ is random.\\
    5) Receiver Initial State of Charge

    State of charge (SOC), i.e., the battery remaining capacity percentage, is equivalent to a fuel gauge for the battery pack in a battery electric vehicle (BEV). That is, the receiver SOC is the percentage of the battery remaining capacity (0\% = empty, 100\% = full). We assume that the initial SOC of each receiver is a random number between 0\% and 100\%. The initial SOC of the receiver $R_i$ is:
    \begin{equation}\label{remaining-capacity}
    \begin{aligned}
    R_{soc}(R_i)=randi([0,100],1,1)\%.
    \end{aligned}
    \end{equation}\\
    6) Charging Time Slot

    For the FAFC scheduling algorithm, the charging time is divided into consecutive equal time slots. The time slot is the minimum charging time unit for the chosen receivers, and it can be set according to the scheduling conditions (e.g., the charging time, the receiver number and so on) in the scheduling process.


    The parameters related to the initial charging status are specified in this subsection. To realize the scheduling process, the core step is the charging and discharging. In next subsections, we will analyze the charging and discharging power for each receiver based on quantitative analysis.

\subsection{Battery Preferred Charging Power Model}\label{}
    To get the battery preferred charging power, we systematically analyze the charging profile of lithium-ion battery in this subsection. Based on analyzing battery charging status, the closed-form formula between the battery preferred charging power and receiver SOC will be obtained.\\
    1) Lithium-ion Battery Charging Profile

\begin{figure}[!t]
	\centering
    \includegraphics[scale=0.67]{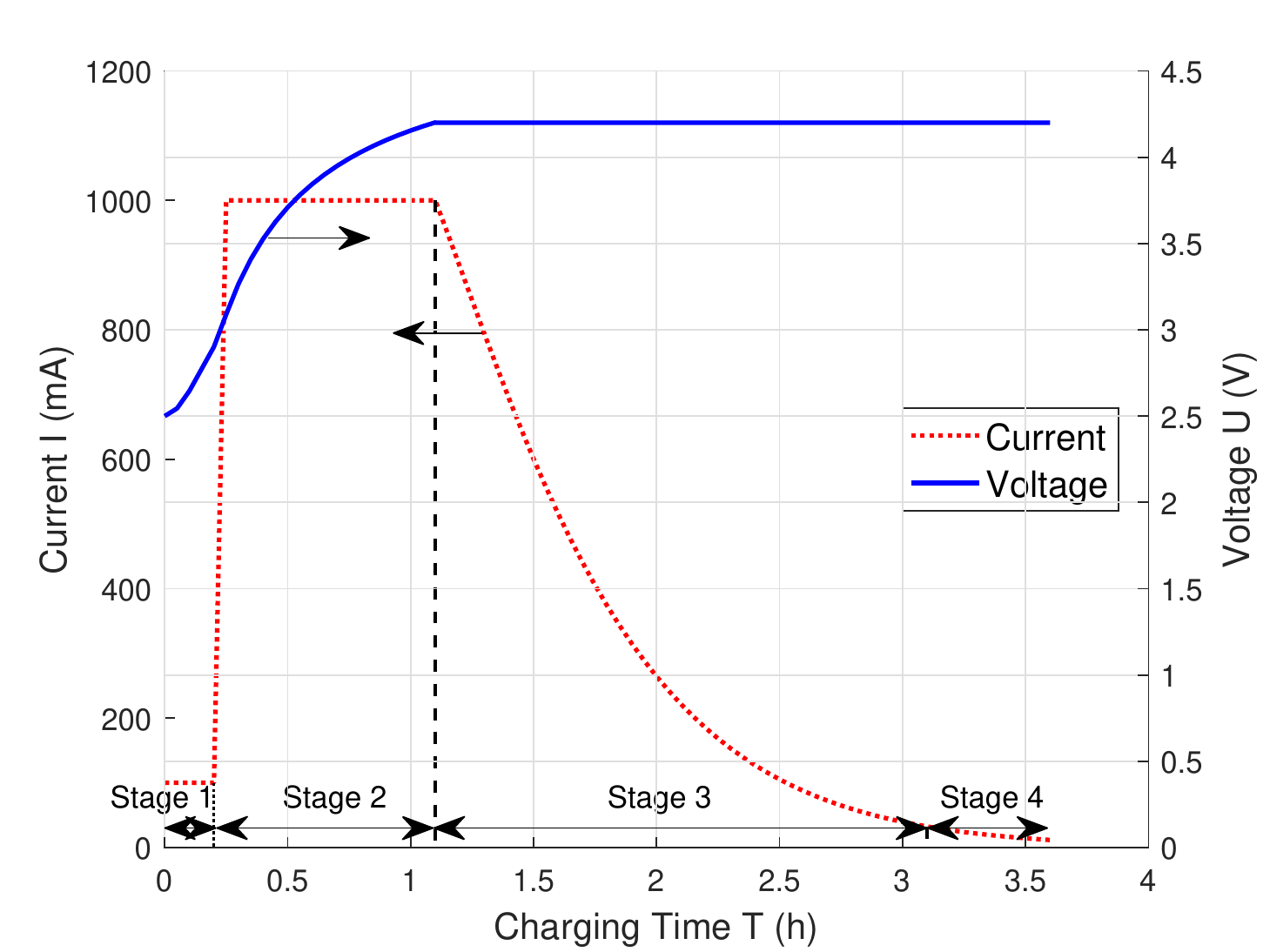}
	\caption{Li-ion Battery Charging Profile}
    \label{chargeproflie}
    \end{figure}

    If batteries are charged with fixed current and voltage, it may cause the undercharging or overcharging problem \cite{dearborn2005charging}. Hence, to optimize the battery charging performance, the constant current-constant voltage (CC-CV) li-ion battery charging profile was researched in \cite{hussein2011review, park2008universal, dearborn2005charging}. The charging profile of 4.2V/1A, 1000mAh lithium-ion battery is shown in Fig. \ref{chargeproflie}.

    In Fig. \ref{chargeproflie}, the charging process of the lithium-ion battery is divided into four stages:

    Stage 1: Trickle Charge (TC) - When the voltage is less than 3V, the lithium-ion battery is charged with a low charging current, which is about 100mA.

    Stage 2: Constant Current (CC) - When the voltage reaches to 3V, the charging current increases from 100mA to 1000mA. Then, the battery is charged with the charging current about 1000mA until the charging voltage increases from 3V to 4.2V.

    Stage 3: Constant Voltage (CV) - At this stage, the battery is charged with a charging voltage about 4.2V, the charging current reduces from 1000mA gradually.

    Stage 4: Charge Termination (CT) - There are two methods to terminate the entire charging process: a) A minimum charging current: when the charging current in the CV stage diminishes to 20mA, the entire charging process terminates. b) A timer: when the CV stage lasts for about 2 hours, the charging procedure terminates. Moreover, to terminate the charging process more precisely, a combination of the two termination methods can be applied.\\
    2) Battery Charging Power and Charging Energy

    From the charging profile of the lithium-ion battery in Fig. \ref{chargeproflie}, the charging current and voltage vary with the charging time. In the charging process, the charging power $P$ of lithium-ion battery can be obtained by multiplying the charging voltage $U$ and current $I$ as:
    \begin{equation}\label{power-VC}
    P = UI.
    \end{equation}

\begin{figure}[!t]
	\centering
    \includegraphics[scale=0.67]{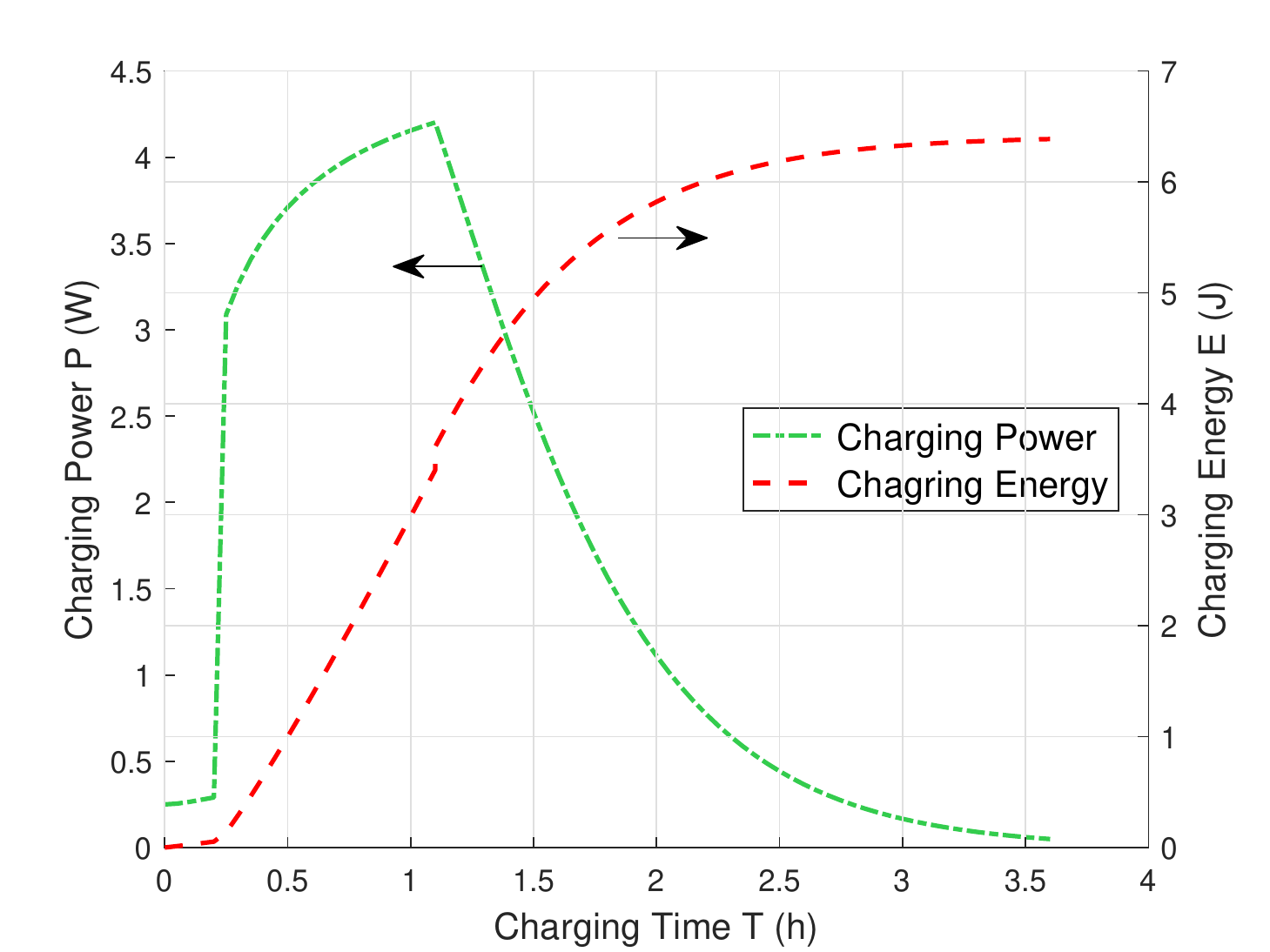}
	\caption{Charging Power and Energy vs. Charging Time}
    \label{profile-power-energy-time}
    \end{figure}

    As the dot-dash curve in Fig. \ref{profile-power-energy-time} shows, the battery is charged with dynamic power during the whole charging procedure. Therefore, the battery charging power is the function of the charging time.

    Moreover, as we all know, the charging energy $E$ is the integral of the charging power over time, as:
    \begin{equation}\label{battery-energy}
    E = \int_{t_{1}}^{t_{2}} P(t)\,dt,
    \end{equation}
    where the charging energy unit is Wh (i.e., J), $t_{1}$ and $t_{2}$ are the lower and upper bounds of the integral time, $P(t)$ is the function of the charging time $t$.

    Based on \eqref{battery-energy}, we can obtain how the charging energy $E$ varies over time from 0 to 3.6h, which is shown as the dash curve in Fig. \ref{profile-power-energy-time}. The battery charging energy starts with a slow growth trend, and then increases gradually as the charging time advances, finally turns to a approximative plateau. Thus, the charging energy also is the function of the charging time. Moreover, the battery total energy $E_o$ is about 6.3865J when the battery is fully charged.\\
    3) Battery Preferred Charging Power and Battery Energy

\begin{figure}[!t]
	 \centering
     \includegraphics[scale=0.67]{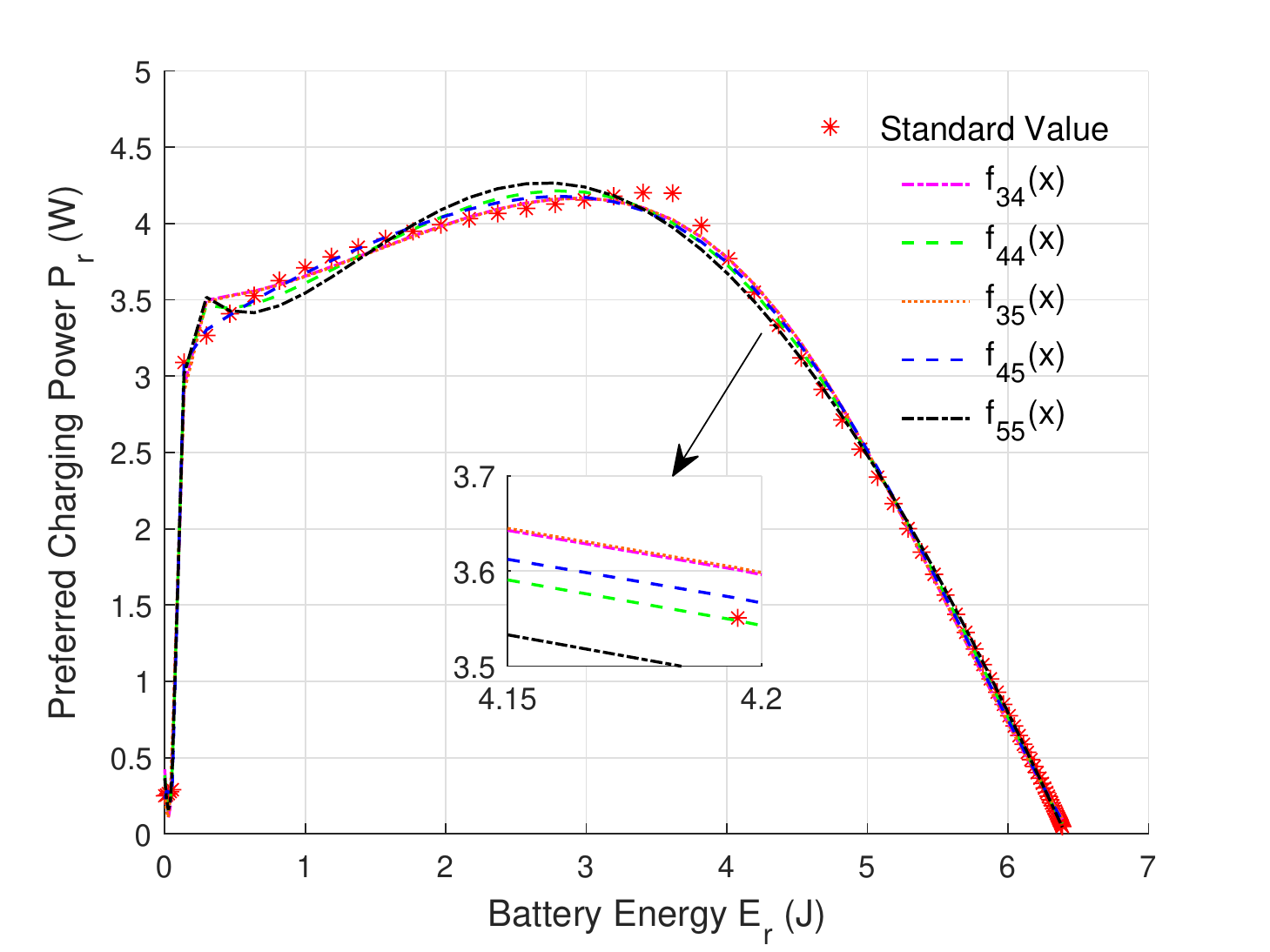}
	 \caption{Preferred Charging Power vs. Battery Energy}
     \label{profile-power-energy-fit}
     \end{figure}

     From Fig. \ref{profile-power-energy-time}, the battery charging power and the charging energy depend on the charging time. The charging time increases gradually according to a fixed step size, and each charging time corresponds to the unique values of the charging power and the battery energy. Therefore, the relationship between the battery preferred charging power $P_r$ and the battery energy $E_r$ can be depicted as the stars in Fig. \ref{profile-power-energy-fit}. Moreover, the value of the preferred charging power corresponding to each battery energy obtained in Fig. \ref{profile-power-energy-time} is defined as     ``Standard Value".

     To implement the scheduling algorithm considering continuously-varying battery charging requirements, we need a closed-form formula to describe the relationship between the battery preferred charging power and the battery energy. Therefore, we fit the relationship between the two by using the MATLAB curve-fitting toolbox. In order to minimize the fitting inaccuracy, we choose the fitting functions with root mean square error (RMSE) less than 0.1 (RMSE $<$ 0.1) among all available fitting functions in the MATLAB curve-fitting toolbox.

     The rational function satisfies the above curve-fitting criterions. A function $f(x)$ is called a rational function if and only if it can be written in the form as:
     \begin{equation}\label{battery-fitting}
     f(x) = \frac{P(x)}{Q(x)} = \frac{\beta_1 x^{n-1}+ \cdots  +\beta_{n-1} x + \beta_n}{\alpha_1 x^{m-1} + \cdots + \alpha_{m-1} x + \alpha_m},
     \end{equation}
     where $P(x)$ and $Q(x)$ are polynomials in $x$, and $Q(x)$ is not the zero polynomial. The relationships between the preferred charging power and battery energy for the five selected fitting functions are shown as curves in Fig. \ref{profile-power-energy-fit}.

     Form Fig. \ref{profile-power-energy-fit}, the preferred charging power rises sharply at first, and then grows slowly, finally decreases gradually when the battery energy increases. In Fig.8, we use $f_{nm}(x)$ to denote the fitting rational function, of which the highest numerator is $n$ power and the highest denominator is $m$ power as in \eqref{battery-fitting}.

     To choose the fitting function which has the better fitting accuracy in Fig. \ref{profile-power-energy-fit}, we calculate the square error $S_{se}$ between the fitting function values $P_{rf}$ and the standard numerical values $P_{rv}$ obtained in Fig. \ref{profile-power-energy-fit} based on \eqref{battery-SE-fitting}, which is:

    \begin{equation}\label{battery-SE-fitting}
    S_{se}=(P_{rf}-P_{rv})^2.
    \end{equation}

    \begin{figure}[!t]
	\centering
    \includegraphics[scale=0.67]{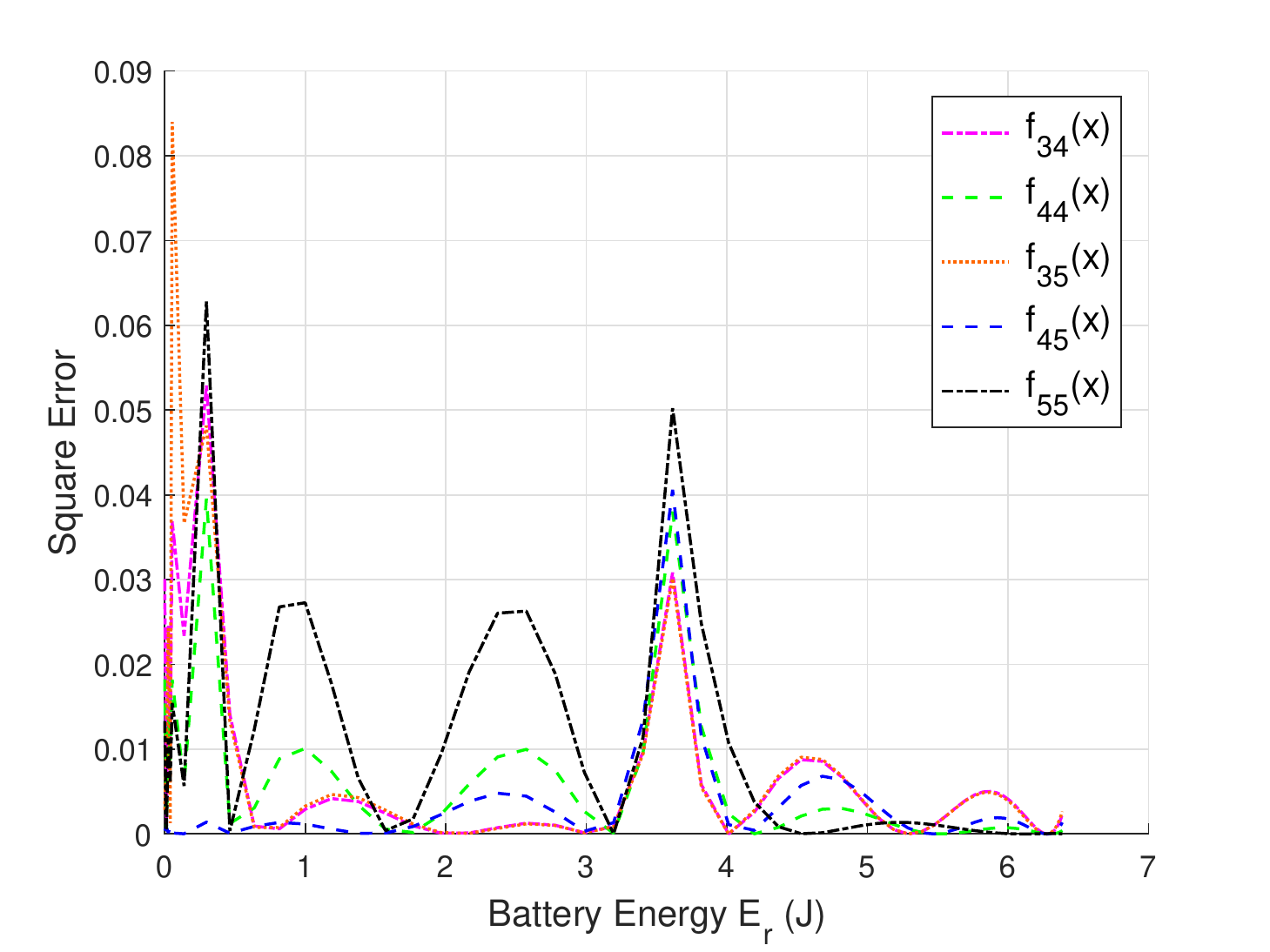}
	\caption{Square Error vs. Battery Energy}
    \label{profile-SE}
    \end{figure}
    
    \begin{figure}[!t]
	\centering
    \includegraphics[scale=0.6]{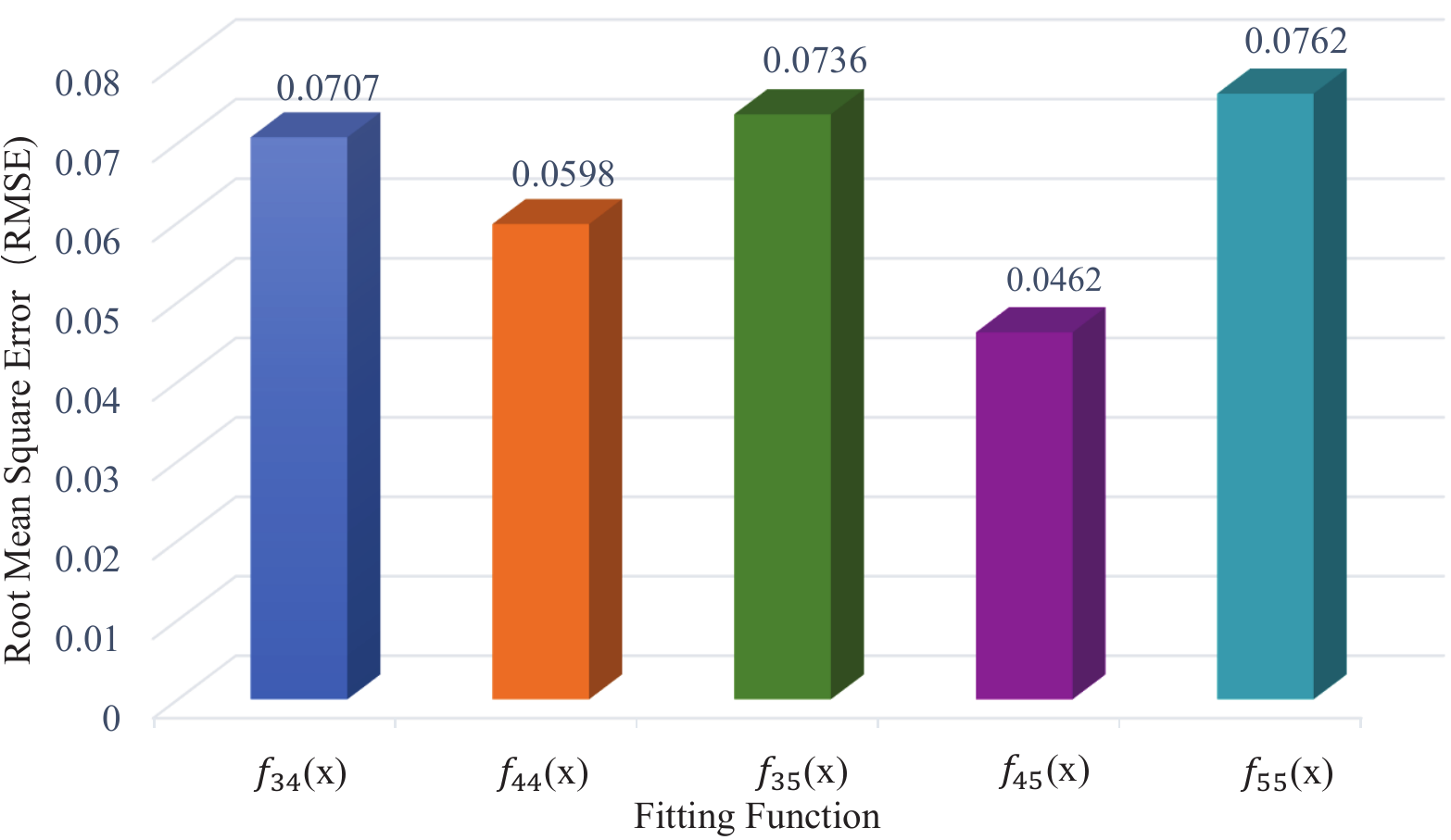}
	\caption{Root Mean Square Error vs. Fitting Function}
    \label{RMSE}
    \end{figure}

     $S_{se}$ of each fitting function in Fig.8 is shown in Fig. \ref{profile-SE}. From Fig. \ref{profile-SE}, the square error of each fitting rational function is small, and the maximum one is about 0.085. To quantize the overall fitting accuracy of the five fitting rational functions in Fig. \ref{profile-power-energy-fit}, we calculate the RMSE $R_{se}$ respectively based on \eqref{RMSE-function}. The RMSE values of all fitting functions are depicted in Fig. \ref{RMSE}.
    \begin{equation}\label{RMSE-function}
     R_{se} = \sqrt{\frac{\sum_{i=1}^n (P_{rf}-P_{rv})^2}{n}}.
     \end{equation}

    From Fig. \ref{RMSE}, the difference of RMSE for the five fitting rational functions is small, and the RMSE of $f_{44}(x)$ and $f_{45}(x)$ are smaller than others. For minimizing the fitting inaccuracy, we choose $f_{44}(x)$ and $f_{45}(x)$ to evaluate the FAFC scheduling algorithm. The two fitting functions are depicted as:

    \begin{itemize}
      \item The function with 4 power numerator and 4 power denominator is:
        \begin{equation}\label{power-soc-fitting44}
        P_{r44}(x) = \frac{\beta_1 x^{4}+ \beta_2 x^{3} + \beta_3 x^{2} +\beta_4 x + \beta_5}{x^{4} + \alpha_1 x^{3} + \alpha_2 x^{2} +\alpha_3 x + \alpha_4}.
        \end{equation}

      \item The function with 4 power numerator and 5 power denominator is:
        \begin{equation}\label{power-soc-fitting45}
        P_{r45}(x) = \frac{{\beta_1}' x^{4}+ {\beta_2}' x^{3} + {\beta_3}' x^{2} + {\beta_4}' x + {\beta_5}' }{x^{5} + {\alpha_1}' x^{4} + {\alpha_2}' x^{3} + {\alpha_3}' x^2 + {\alpha_4}'x +{\alpha_5}'}.
        \end{equation}
    \end{itemize}

    In \eqref{power-soc-fitting44} and \eqref{power-soc-fitting45}, $P_{r44}(x)$ and $P_{r45}(x)$ are the preferred charging power calculated by the two fitting functions, and $x$ is the battery energy $E_r$. The values of all the coefficients in \eqref{power-soc-fitting44} and \eqref{power-soc-fitting45} are shown in Table \ref{Coefficient}.
    
    \begin{figure}[!t]
	\centering
    \includegraphics[scale=0.67]{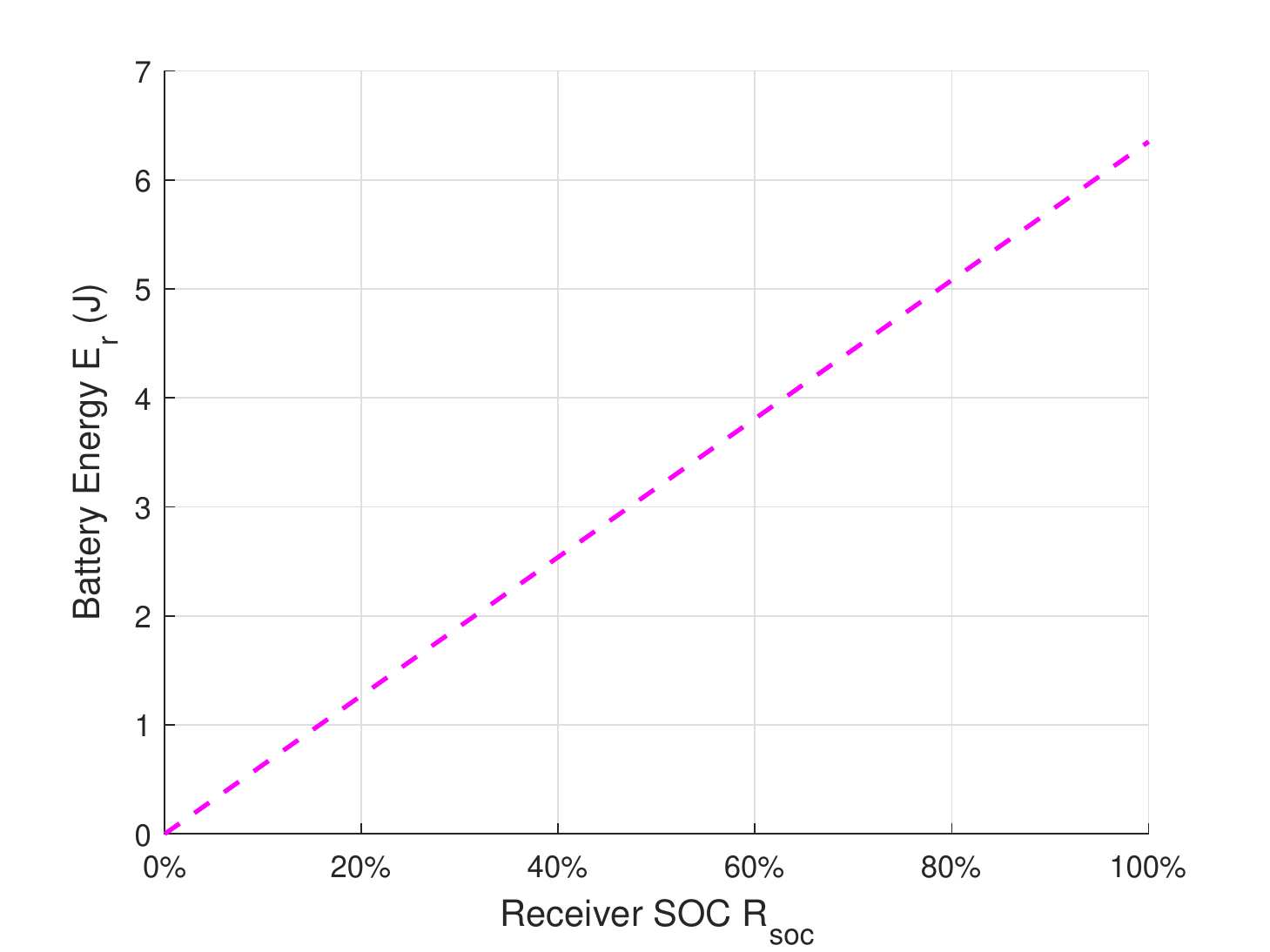}
	\caption{Battery Energy vs. Receiver SOC}
    \label{profile-energy-soc}
    \end{figure}
    \begin{table}[!t]
    \centering
    \caption{Fitting Functions Coefficients}
    \begin{tabular}{C{1.5cm} C{2cm} |C{1.5cm} C{2cm}}
    \hline
    \multicolumn{2}{c|}{$P_{r44}$} & \multicolumn{2}{c}{$P_{r45}$}\\
    \textbf{Coefficient} & \textbf{Value} & \textbf{Coefficient} & \textbf{Value}\\
    \hline
    $\beta_1$ & {$-3.112$} & ${\beta_1}'$ & {$-21.65$}\\
    $\beta_2$ & {$1.439$} &  ${\beta_2}'$ & {$141.2$}\\
    \bfseries{$\beta_3$} & {$120.4$}  &{${\beta_3}'$} & {$-11.5$} \\
    \bfseries{$\beta_4$} & {$-7.452$} &{${\beta_4}'$} & {$0.1526$}\\
    \bfseries{$\beta_5$} & {$0.1543$} &{${\beta_5}'$} & {$0.008358$}\\
    \bfseries{$\alpha_1$} & {$-9.881$} &{${\alpha_1}'$} & {$-10.7$}\\
    \bfseries{$\alpha_2$} & {$44.84$} &{${\alpha_2}'$} & {$41.01$}\\
    \bfseries{$\alpha_3$} & {$-5.49$} &{${\alpha_3}'$} & {$ -1.509$}\\
    \bfseries{$\alpha_4$} & {$ 0.4007$} &{${\alpha_4}'$} & {$-0.3997$}\\
    \bfseries{$\  $} & {$\  $} &{${\alpha_5}'$} & {$0.0362$}\\
    \hline
    \label{Coefficient}
    \end{tabular}
    \end{table}

    \noindent 4) Battery Preferred Charging Power and Receiver SOC

    In addition, the receiver SOC $R_{soc}$ is the ratio of the battery energy $E_r$ to the total energy $E_o$:
    \begin{equation}\label{battery-energy-soc}
    R_{soc} = \frac{E_r}{E_o} \times 100\%.
    \end{equation}

    From \eqref{battery-energy-soc}, given the total energy $E_o$, the battery energy $E_r$ has a linear relationship with the receiver SOC $R_{soc}$, which is depicted in Fig. \ref{profile-energy-soc}. Thus, the battery energy $E_r$ can be obtained from the receiver SOC $R_{soc}$. For example, $E_r$ is 3.8319J when $R_{soc}$ is 60\%.

\begin{figure}[!t]
	\centering
    \includegraphics[scale=0.67]{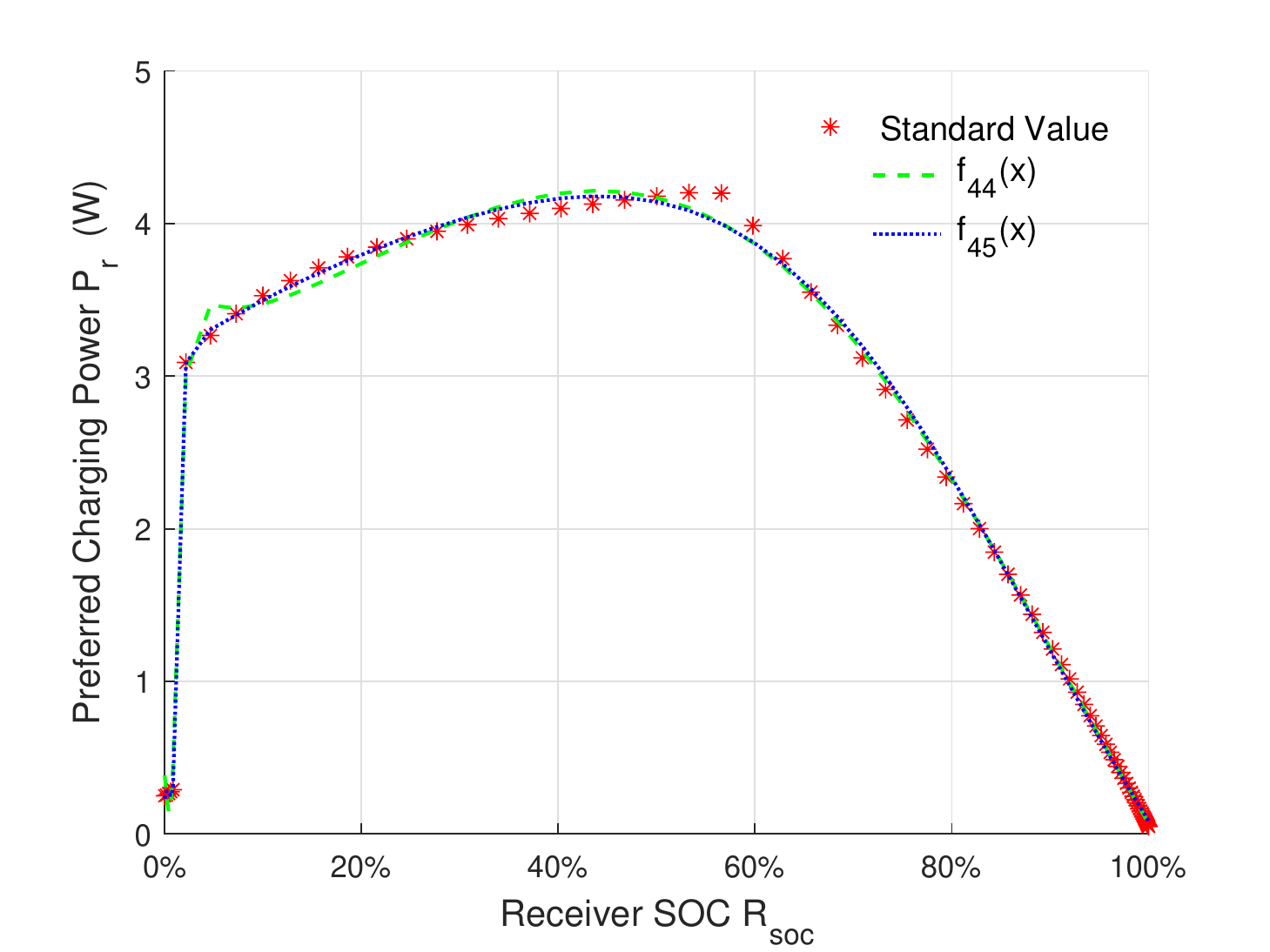}
	\caption{Preferred Charging Power vs. Receiver SOC}
    \label{profile-chargepower-soc}
    \end{figure}

    From Figs. \ref{profile-power-energy-fit} and \ref{profile-energy-soc}, we can obtain the relationship between the receiver SOC and the preferred charging power for the fitting functions $f_{44}(x)$ and $f_{45}(x)$, which are depicted in Fig. \ref{profile-chargepower-soc}.

    From formulas \eqref{power-soc-fitting44}, \eqref{power-soc-fitting45} and \eqref{battery-energy-soc}, the receiver preferred charging power $P_r$ can be obtained from the receiver SOC $R_{soc}$. For example, the SOC of receiver $R_i$ is 60\%, from \eqref{power-soc-fitting44} and \eqref{battery-energy-soc}, the preferred charging power $P_{r44}$ is 3.8650W. Based on \eqref{power-soc-fitting45} and \eqref{battery-energy-soc}, the preferred charging power $P_{r45}$ is 3.8712W.
%
\subsection{Battery Discharging Power Model}\label{}
    \begin{figure}[!t]
	\centering
    \includegraphics[scale=0.5]{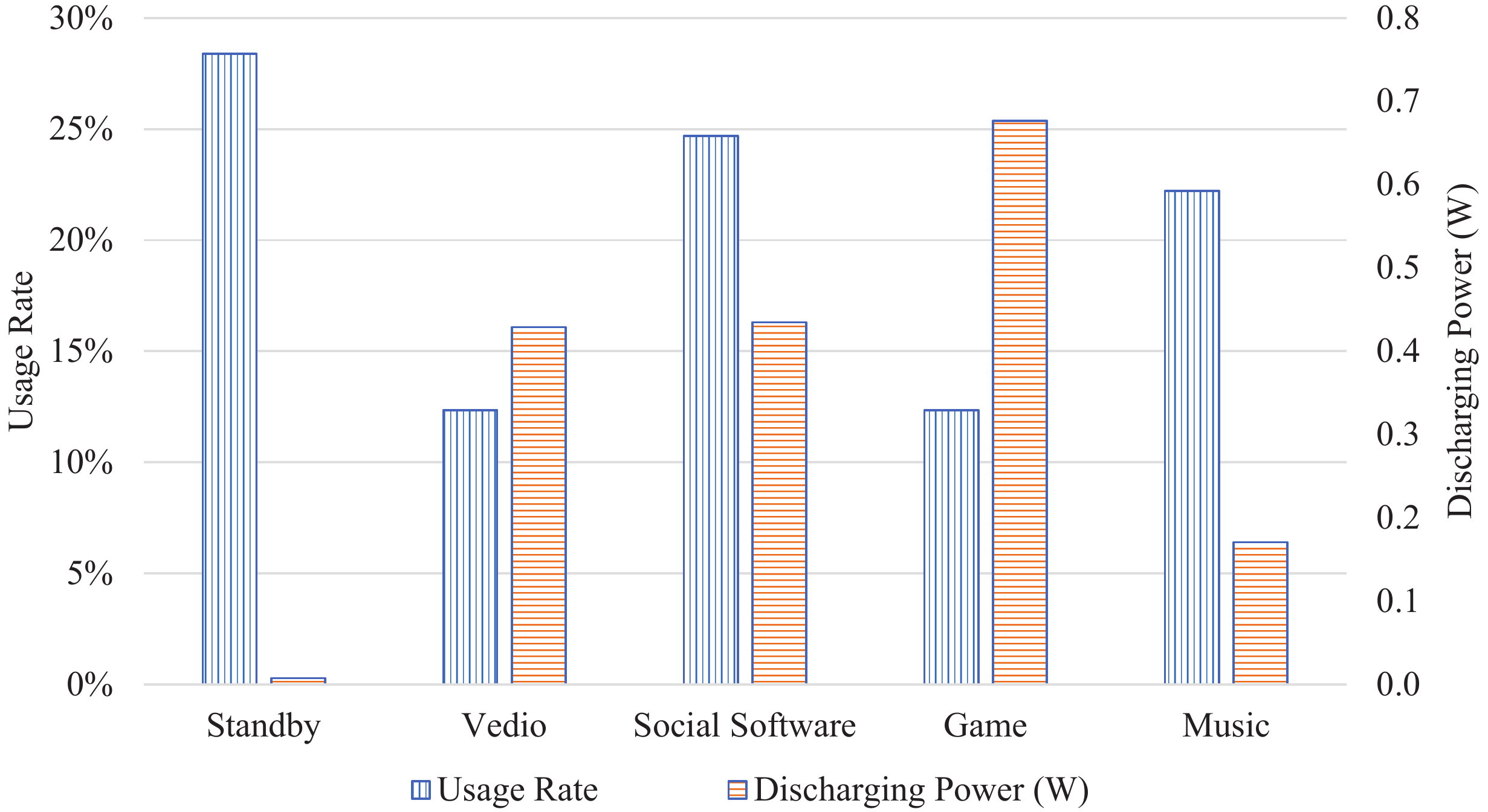}
	\caption{Usage Rate and Discharging Power vs. Working Status}
    \label{dischargepower}
    \end{figure}

    The discharging power of each receiver depends on the battery type and the working status \cite{carroll2010analysis, peng2014smartphone, murmuria2012mobile}. For example, the receiver discharging power of lithium-ion battery is different from that of Ni-MH battery, and playing music is different from playing video with the same battery. To analyze the discharging power quantitatively in the scheduling algorithm, we propose the discharging power model for the receivers with different working status.

    Based on the survey of the mobile phone applications, the representative working status of mobile phone include: standby, video, social software (e.g., Facebook, Twitter), game and music \cite{Usingfrequency}. We assume that a receiver being charged works at one of the above five working status. The usage rate and discharging power of the five working status are depicted in Fig. \ref{dischargepower}.

    In Fig. \ref{dischargepower}, the vertical strip column is the usage rate of each working status, while the horizontal strip column is the discharging power. From Fig. \ref{dischargepower}, the discharging power of playing games is maximal, and standby is the most common working state.

    Therefore, during a charging time slot, the discharging power of a receiver can be modeled as:

    \begin{itemize}
      \item Discharging power of each working status:
      \begin{equation}\label{powerconsumption}
        \begin{aligned}
        P_{u}=\{0.0076\  0.4289\  0.4348\  0.6766\  0.1706\}.
         \end{aligned}
        \end{equation}
      \item Usage rate of each working status:
        \begin{equation}\label{probability}
        \begin{aligned}
        U_{p}=\{28.39\%\  12.35\%\  24.69\%\  12.35\%\  22.22\%\}.
        \end{aligned}
        \end{equation}
      \item Discharging power of a receiver during a time slot is:
        \begin{equation}\label{}
        \begin{aligned}
        P_{d}=randsrc(1,1,[P_{u};U_{p}]).
        \end{aligned}
        \end{equation}
    \end{itemize}

    In summary, all parameters related to the algorithm are quantified in the above contents, so we can implement the algorithm based on the algorithm principle presented in Section III and the quantized parameters analyzed in this section. The operational pseudo code of FAFC scheduling algorithm will be presented in the next subsection.
    \begin{algorithm}
    \caption{FAFC Scheduling Algorithm}
    \begin{algorithmic}[1]
    \Require $N_r$, $R_{soc}$;
    \State initialize $T_{c}$, $T_{p}$, $T_{s} \leftarrow 0$, $i \leftarrow 1$, $flag \leftarrow 0$;
    \State $E_r \leftarrow R_{soc}\times E_o$;
    \While {$E_r \geqslant 0$ \textbf{and} $E_r < E_o$ \textbf{and} $T_s \leqslant T_p$}
    \State initialize $P_t$;
    \State $P_o \leftarrow P_t$;
    \State $P_d \leftarrow randsrc(1,N_r,[P_u;U_p])$;
    \State $P_d \leftarrow [P_d(flag:end)\  P_d(1:flag)]$;
    \State do \eqref{power-soc-fitting44} (or \eqref{power-soc-fitting45}) to get $P_{r}$;
    \For {$i \leftarrow 1\ to\ N_r$}
    \If {$P_o \geqslant P_r(i)$}
    \State $P_c(i)\leftarrow P_r(i)$;
    \State $E_r(i)\leftarrow E_r(i)+P_c(i)\times T_c-P_d(i)\times T_c$;
    \State $P_o \leftarrow P_o-P_c(i)$;
    \If {$P_o = 0$} $flag \leftarrow i$;
    \EndIf
    \ElsIf {$P_o > 0$}
    \State $P_c(i) \leftarrow P_o$;
    \State $E_r(i) \leftarrow E_r(i)+P_c(i)\times T_c-P_d(i)\times T_c$;
    \State $P_o \leftarrow P_o-P_c(i)$;
    \If {$P_o=0$} $flag \leftarrow i$;
    \EndIf
    \Else
    \State $E_r(i) \leftarrow E_r(i)-P_d(i)\times T_c$;
    \EndIf
    \EndFor
    \State $E_r \leftarrow [E_r(flag:end)\    E_r(1:flag)];$
    \State $T_s \leftarrow T_s + T_c$;
    \EndWhile
    \State \Return{$\frac{E_r}{E_o} \times 100\%$};
    \label{FAFCA}
    \end{algorithmic}
    \end{algorithm}
\subsection{Operational Pseudo Code}\label{}

    Based on the algorithm principle and the quantized parameters, the operational pseudo code of the FAFC scheduling algorithm can be written as Algorithm $1$. The parameters of the algorithm are summarized in Table \ref{Parameters of model}.

    Based on the FAFC scheduling algorithm, given the receiver number and the initial SOC, the receivers, which discharge at different power, can be charged chronologically with their preferred charging power according to the accessing time. Keeping selecting and charging receivers in the charging process, the transmitting power is scheduled to keep all receivers working as long as possible for fairness. In the charging process, we record the receiver SOC of each receiver and the charging time to determine when to terminate the scheduling process.

    \begin{table}[!t]
    \centering
        \caption{FAFC Execution Scheduling Parameter}
    \begin{tabular}{C{1.5cm} C{6cm}}
    \hline
     \textbf{Symbol} & \textbf{Parameter}  \\
    \hline
    \bfseries{${N}_{r}$} & {Receiver number} \\
    \bfseries{${R}_{soc}$} & {Receiver SOC (remaining capacity percentage)} \\
    \bfseries{${T}_{c}$} & {Charging time slot} \\
    \bfseries{${T}_{p}$} & {Pre-set charging time} \\
    \bfseries{${T}_{s}$} & {Scheduling process charging time} \\
    \bfseries{${E}_{r}$} & {Battery energy} \\
    \bfseries{${E}_{o}$} & {Battery total energy} \\
    \bfseries{${P}_{o}$} & {Available transmitting power} \\
    \bfseries{${P}_{u}$} & {Discharging power of receiver working status}\\
    \bfseries{${U}_{p}$} & {Usage rate of receiver working status}\\
    \bfseries{${P}_{d}$} & {Receiver discharging power}\\
    \bfseries{${P}_{r}$} & {Receiver preferred charging power}\\
    \hline
    \label{Parameters of model}
    \end{tabular}
    \end{table}

\section{Performance Analysis}
     The performance analysis of the FAFC scheduling algorithm is based on the simulation in MATLAB. Since the receiver can keep working when its SOC is not zero, the SOC of each receiver reflects the algorithm's performance. The impacts of the receiver number, the charging time, and the transmitting power on the receivers' SOC will be evaluated in this section. The principles of simultaneous charging and the methods to improve the charging performance in the multi-user RBC system will be obtained finally.

\subsection{Simulation Design}\label{}
    The simulation is divided into two parts: 1) the impacts that the charging time and the receiver number have on the receivers' SOC; 2) the impacts that the transmitting power and the receiver number have on the receivers' SOC. The parameters of the simulation are illustrated as follows.

    For 1), the receiver number $N_r$ varies from 10 to 50, while the transmitting power $P_t$ is 20W. We compare the receivers' average SOC after the receivers being charged 1, 2, 3 hours by the transmitter, and analyze the impacts of the receiver number and the charging time on the receivers' SOC.

    For 2), the receiver number $N_r$ varies from 10 to 50, the charging time T is 3 hours, and the transmitting power $P_t$ takes 20W, 40W, 60W, 80W and 100W, respectively. After the charging process, the receivers' average SOC under different transmitting power is compared. The impacts of the transmitting power and the receiver number on the receivers' SOC are illustrated by the comparison.

    Moreover, we set the time slot $T_c$ as 10s by calculating the ratio of the charging time with the battery capacity. Since the parameters (receiver initial SOC, discharging power, etc.) of each receiver are random during the scheduling process,  the ``averaging multiple experiments" method is adopted to eliminate randomness. Therefore, the SOC of each receiver in the simulation results is the result of averaging multiple scheduling simulations.
\subsection{Simulation Results}\label{}
\begin{figure}[!t]
	\centering
    \includegraphics[scale=0.67]{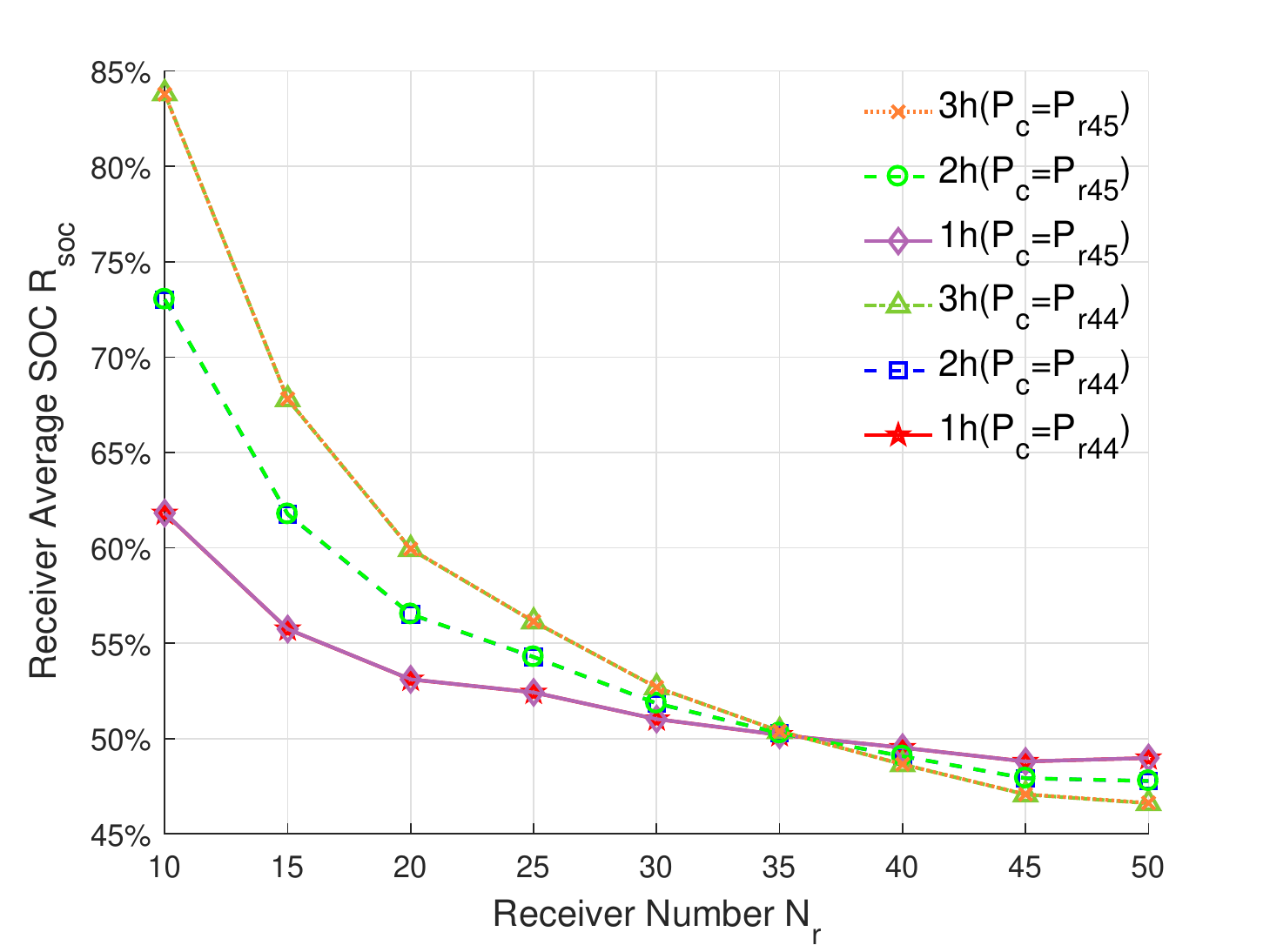}
	\caption{Receiver Average SOC vs. Receiver Number ($P_t$=20W)}
    \label{FAFCtime-com}
    \end{figure}

     \emph{1)} When the transmitting power $P_t$ is 20W, the average SOC of all receivers after being charged for 1, 2, 3 hours is compared. The receiver charging power $P_c$ in a charging time slot is equal to $P_{r44}$ calculated by \eqref{power-soc-fitting44} or $P_{r45}$ calculated by \eqref{power-soc-fitting45}. The charging time 1, 2, 3 hours is the different charging stages of a same charging process. The comparison results are depicted in Fig. \ref{FAFCtime-com}.

    In Fig. \ref{FAFCtime-com}, each curve denotes the average SOC for different receiver number and different charging time. The receiver SOC decreases as the receiver number increases since the total consumed energy increases as the receiver number increases, while the total charging energy is fixed in each time slot. Moreover, when the charging energy is greater than the consumed energy, the receiver SOC increases as the charging time prolongs, e.g., when the receiver number is less than 35. Rather, the average SOC decreases when the receiver number is larger than 35.

\begin{figure}[!t]
	\centering
    \includegraphics[scale=0.67]{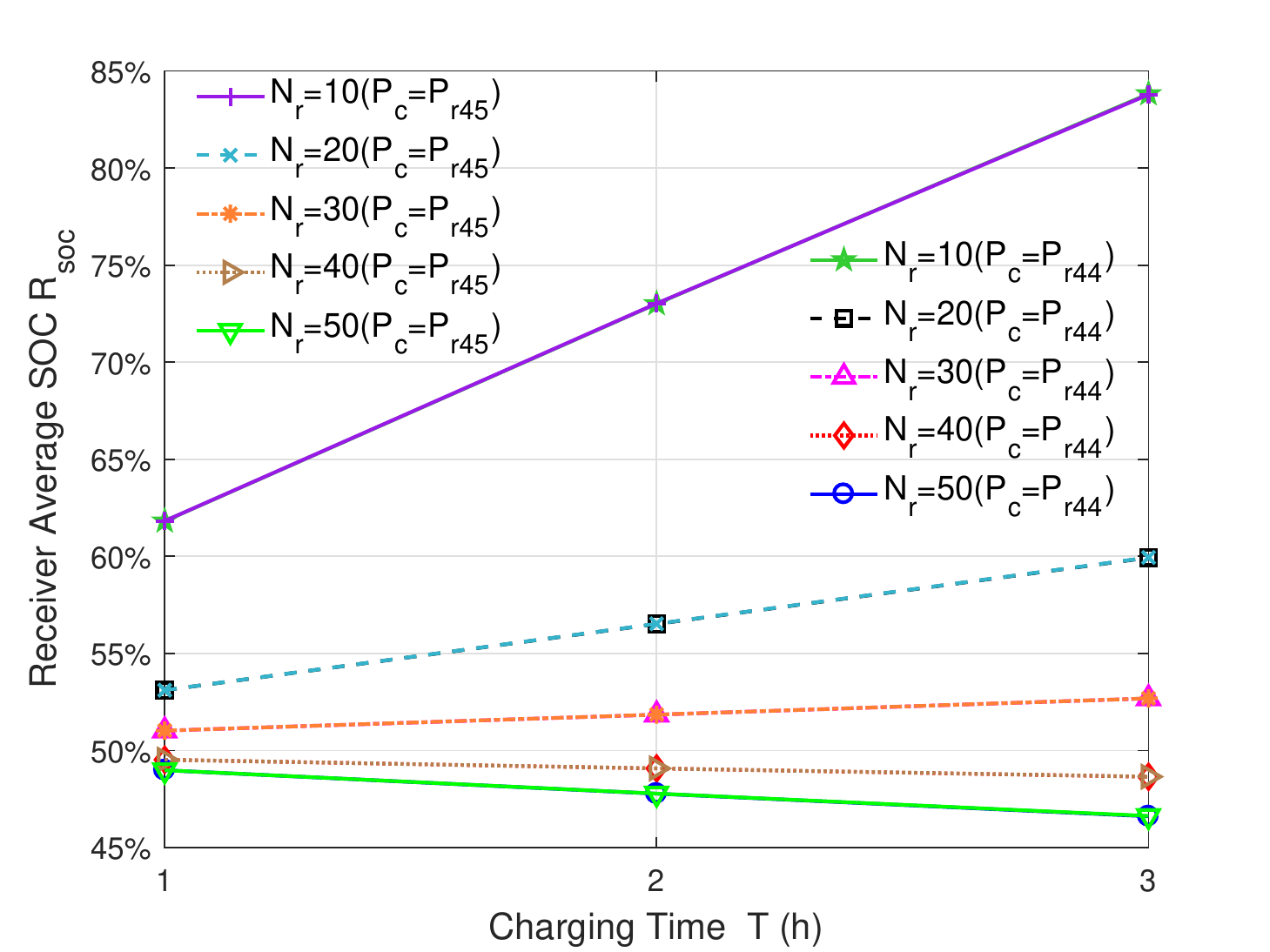}
	\caption{Receiver Average SOC vs. Charging time ($P_t$=20W)}
    \label{FAFCtime-num}
    \end{figure}

    Given the different charing time, to illustrates the impacts of the receiver number on the average SOC, we show the variation of the average SOC with different receiver numbers in Fig. \ref{FAFCtime-num}.
        \begin{figure}[!t]
	\centering
    \includegraphics[scale=0.67]{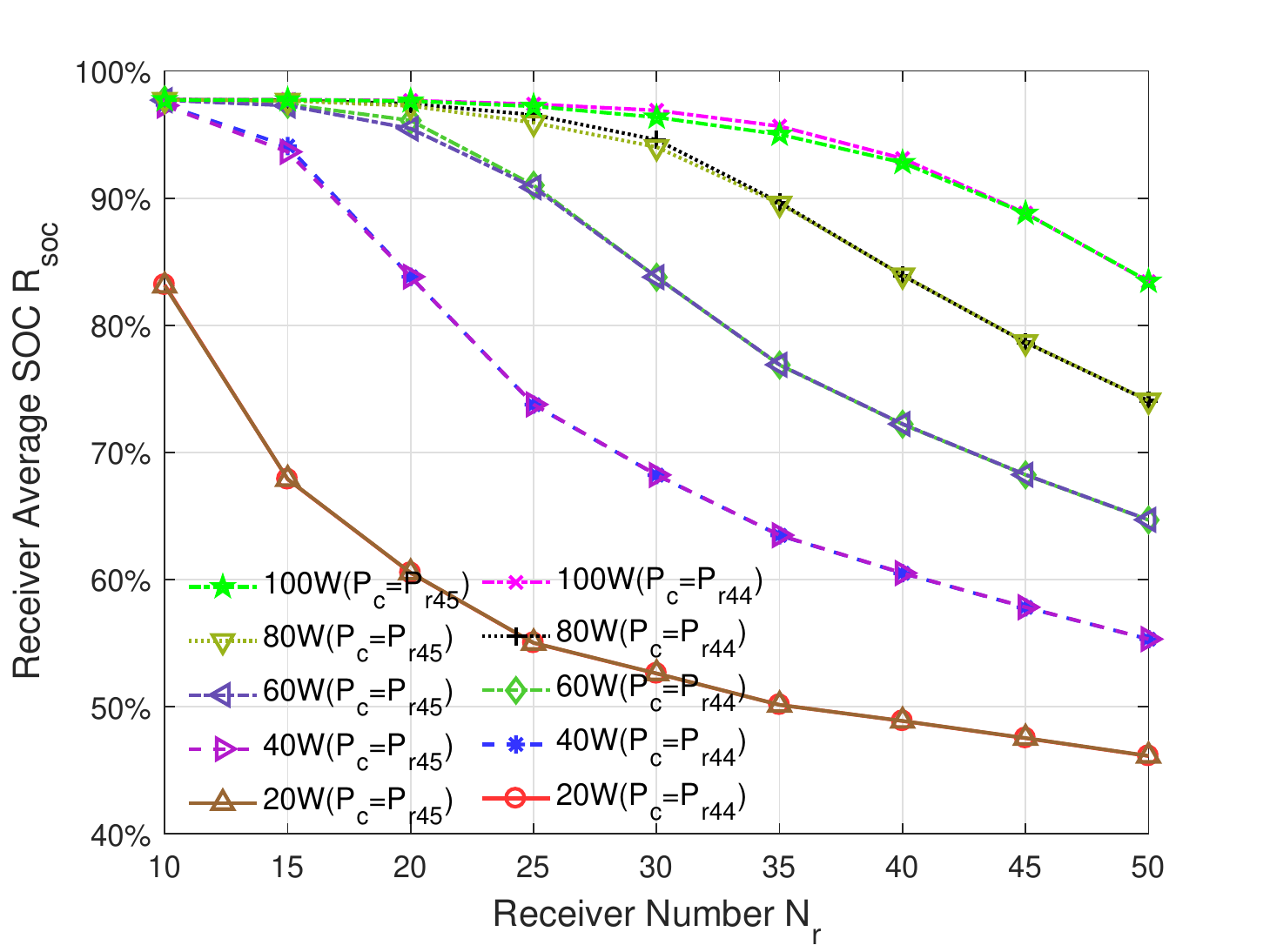}
	\caption{Receiver Average SOC vs. Receiver Number (T=3h)}
    \label{FAFCpower-com}
    \end{figure}

    In Fig. \ref{FAFCtime-num}, the receiver SOC increases with the charging time prolonging when the receiver number is small, since the consumed energy is less than the charging energy. As the receiver number is about 30, the receiver SOC is close to steady regardless of the charging time changing. Moreover, the receiver SOC decreases as the charging time prolongs when the receiver number $N_r$ is 40 and 50.

    Thus, to maximize the receiver SOC, the charging time should be prolonged when the charging energy is greater than the consumed energy. Furthermore, when the transmitting power is fixed, there exists a threshold for the number of receivers being charged simultaneously, in order to avoid the system running out of power.

    \emph{2)} To illustrate the impacts of the transmitting power on the receiver SOC, we depict the receivers' average SOC in Fig. \ref{FAFCpower-com} after receivers being charged for 3 hours with 20W, 40W, 60W, 80W and 100W transmitting power.

    In Fig. \ref{FAFCpower-com}, each curve denotes the average SOC for different receiver number and transmitting power. When the transmitting power is fixed, the average SOC decreases gradually since the consumed energy increases as the receiver number grows. Moreover, when the receiver number is same, the receiver SOC increases with the transmitting power increasing.

\begin{figure}[!t]
	\centering
    \includegraphics[scale=0.67]{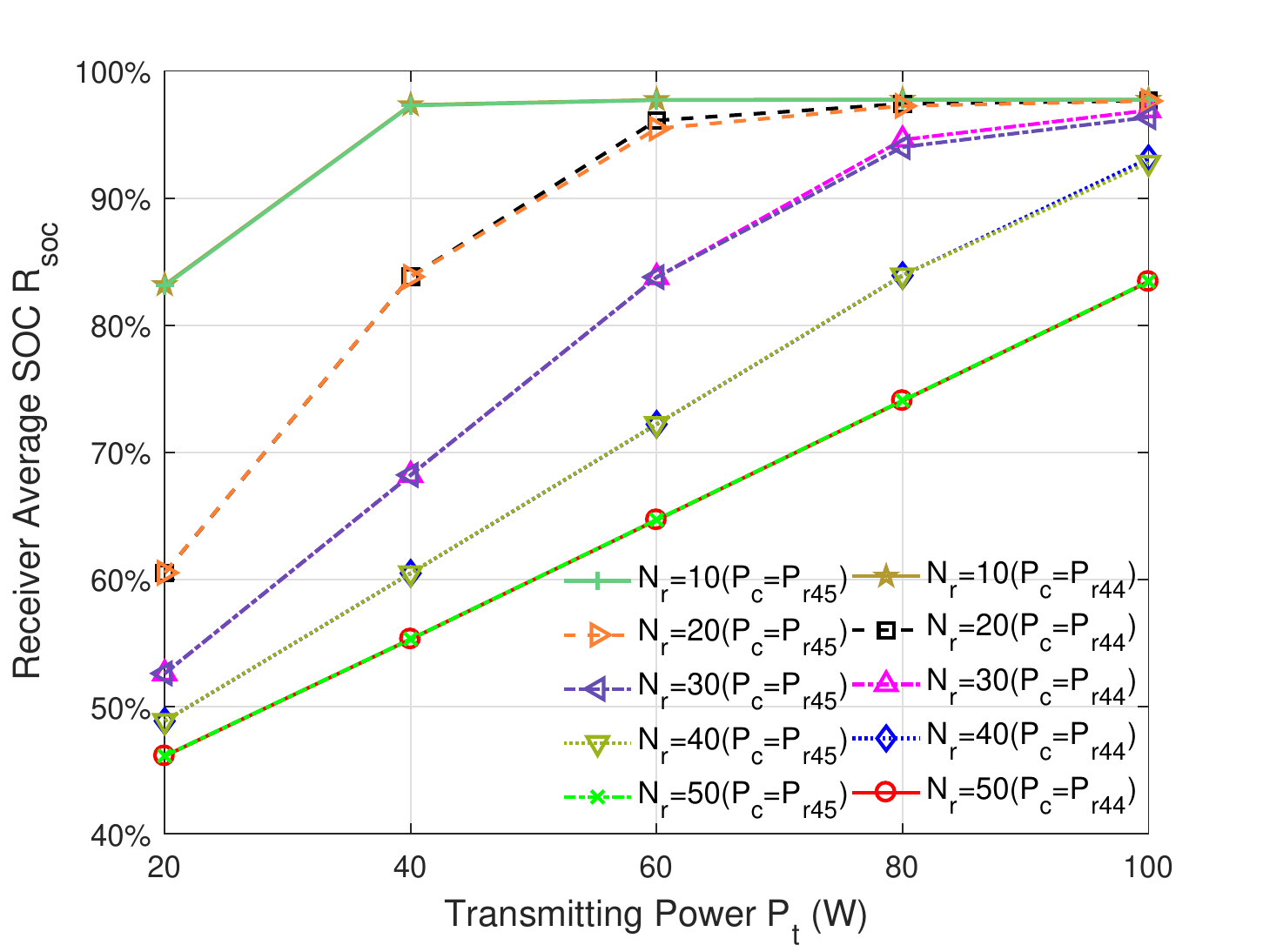}
	\caption{Receiver Average SOC vs. Transmitting Power (T=3h)}
    \label{FAFCpower-num}
    \end{figure}

    To explain the impacts of the receiver number $N_r$ under different transmitting power on the receiver SOC, the variation trend of the average receiver SOC with different receiver numbers is shown in Fig. \ref{FAFCpower-num}.

    From Fig. \ref{FAFCpower-num}, for the same receiver number, the receiver SOC rises when the transmitting power increases. When the receiver number is small and the transmitting power is high, since the charging energy can satisfy the consumed energy, the receivers can be fully charged approximately. For example, when the receiver number is 10, all the receivers can be almost fully charged if the transmitting power is greater than 40W.

    Therefore, if the transmitter power source can meet the charging requirements, the transmitting power should be improved to extend the working time of all receivers. Moreover, the charging power for receivers will increase as the charing efficiency raises. For example, when the source power $P_s$ is 50W, the charging power $P_e$ is 10W with the charing efficiency ${\eta}_{o}$ of 20\%, while $P_e$ is 15W with 30\% charing efficiency ${\eta}_{o}$. Thus, to improve the charging performance, it is also effective to increase the single user's charging efficiency in the multi-user RBC system.

    Moreover, from the Figs. \ref{FAFCtime-com}, \ref{FAFCtime-num}, \ref{FAFCpower-com}, and \ref{FAFCpower-num}, given the charging power $P_{r44}$ and $P_{r45}$, the change trend of the receiver SOC is same and the difference between the two SOC is small. Therefore, the small difference between the two fitting functions has no effect on the simulation results.

\subsection{Analysis Summary}\label{}
    The factors impacts the performance of FAFC scheduling algorithm include: the transmitting power, the receiver number, the charging time and the charging efficiency. The following conclusions are obtained through the simulation results:

    1) When the transmitting power and the receiver number are fixed, if the fixed charging energy is greater than the consumed energy in a charging time slot, the receiver SOC increases with the charging time increasing. Otherwise, the receiver SOC decreases as the charging time prolongs.

    2) The receiver SOC decreases as the receiver number increases, under the fixed transmitting power and the fixed charging time.

    3) Regardless of the variation of the charging time and the receiver number, the receiver SOC grows with the transmitter transmitting power increasing.

    4) When the charging time is 3 hours and the receiver number is 10, the receivers will be almost fully charged when the transmitter transmitting power is greater than 40W.

    To improve the FAFC scheduling algorithm performance (i.e., to keep all receivers working as long as possible for fairness in the multi-user RBC system), the strategies are as follows:

     1) To maximize the SOC of each receiver, the charging time should be prolonged when the charging energy is greater than the consumed energy.

     2) When the transmitting power is fixed, there exists a threshold for the number of receivers being charged simultaneously in order to avoid the system running out of power.

     3) If the power supply can meet the charging requirements, the transmitting power should be improved to extend the working time of all receivers.

     4) For the multi-user RBC system, increasing the charging efficiency of the single user is the essential method to improve the charging performance.

\section{Conclusions}\label{Section5}

     We present the First Access First Charge (FAFC) scheduling algorithm for wireless power transfer in the multi-user RBC system to keep all IoT devices working as long as possible for fairness. The muti-user RBC system includes a transmitter with the fixed transmitting power and multiple receivers working with different battery status and power consumption patterns. Based on the FAFC scheduling algorithm, the transmitting power is scheduled to charge multiple IoT devices according to their accessing time sequence. We quantify the system parameters for the FAFC scheduling algorithm implementation. The receiver preferred charging power can be obtained from its SOC, while the discharging power is determined by the working status of the IoT device. Then, the operational pseudo code is presented for the algorithm implementation. Finally, the simulation demonstrates the performance of FAFC the scheduling algorithm for the multi-user RBC system with the impacts of the transmitting power, the receiver number, and the charging time.

    Based on the performance analysis, we find that, to keep all receivers working as long as possible for fairness, the charging time should be prolonged when the charging energy is greater than the consumed energy. Moreover, the transmitting power should be increased if the transmitter power source can meet the charging demands. Furthermore, the receiver number being charged simultaneously should be limited by a threshold, and the charging efficiency of the single IoT device should be raised to improve the charging performance.

    However, there are still several open issues can be studied in the future, for example:

    \begin{itemize}
      \item In the charging process, the charging urgency of each receiver is more essential, so the scheduling algorithm should be designed considering the charging urgency rather than the accessing time sequence to improve the charging quality of each IoT device.
      \item Throughout the charging process, the charging connections established may be disconnected, and the other receivers may access to the transmitter, which results in the dynamic change of the receiver number. Thus, it is necessary to analyze the performance of the scheduling algorithm when the receiver number is varying.
      \item The scheduling process of multi-transmitter to multi-user could be further investigated.
    \end{itemize}

\bibliographystyle{IEEEtran}
\bibliographystyle{unsrt}
\bibliography{references}

\end{document}